\documentclass[final,1p,times,authoryear]{elsarticle}

\usepackage{amssymb}
\usepackage{amsmath}
\usepackage{amsthm}
\usepackage{mathrsfs}

\usepackage{algorithm}
\usepackage[noend]{algpseudocode}
\usepackage{easyReview}

\begin{document}

\begin{frontmatter}

\title{Scaling global tractography with ExaTract: from large cohorts to ultra-high-resolution datasets} %% Article title

\author[label1,label2]{Simon Legeay} %% Author name
\author[label3,label2b]{Felix Matuschke} %% Author name
\author[label1,label5]{Bastien Herlin} %% Author name
\author[label3,label2b]{Jan André Reuter} %% Author name
\author[label6,label7]{Christophe Destrieux}
\author[label3,label2b,label4]{Markus Axer} %% Author name
\author[label1,label2]{Ivy Uszynski} %% Author name
\author[label1,label2]{Cyril Poupon} %% Author name

%% Author affiliation
\affiliation[label1]{organization={Université Paris-Saclay, CEA, CNRS, Neurospin, UMR 9027 BAOBAB, GAIA Laboratory},%Department and Organization
            addressline={Centre d'études Paris-Saclay, bat 135}, 
            city={Gif-sur-Yvette},
            postcode={91191},
            country={France}}

%% Author affiliation
\affiliation[label2]{organization={AIDAS},%Department and Organization
            city={Gif-sur-Yvette},
            postcode={91191},
            country={France}}

%% Author affiliation
\affiliation[label3]{organization={Institute of Neuroscience and Medicine (INM-1)},%Department and Organization
            addressline={Forschungszentrum Jülich GmbH, Wilhelm-Johnen-Straße}, 
            city={Jülich},
            postcode={52428},
            country={Germany}}

%% Author affiliation
\affiliation[label2b]{organization={AIDAS},%Department and Organization
            city={Jülich},
            postcode={52428},
            country={Germany}}

%% Author affiliation
\affiliation[label5]{organization={Sorbonne University, Frontlab, CNRS UMR 7225, INSERM U1127},%Department and Organization
            addressline={Hôpital Pitié-Salpêtrière, 47 bd de l'Hôpital}, 
            city={Paris},
            postcode={75013},
            country={France}}

%% Author affiliation
\affiliation[label6]{organization={University of Tours, INSERM, Imaging Brain \& Neuropsychiatry iBraiN U1253},%Department and Organization
            addressline={10 boulevard Tonnellé}, 
            city={Tours},
            postcode={37032},
            country={France}}

%% Author affiliation
\affiliation[label7]{organization={CHRU de Tours},%Department and Organization
            addressline={2 boulevard Tonnellé}, 
            city={Tours},
            postcode={37044},
            country={France}}

%% Author affiliation
\affiliation[label4]{organization={School of Mathematics and Natural Sciences, University of Wuppertal},%Department and Organization
            addressline={Gaussstraße}, 
            city={Wuppertal},
            postcode={42119},
            country={Germany}}

%% Abstract
\begin{abstract}
%% Text of abstract
Global tractography offers a theoretical framework for reconstructing white matter connectivity from diffusion MRI by jointly inferring all streamlines and their consistency with the diffusion signal and underlying anatomical priors. However, its sequential nature and log-linear complexity make it computationally prohibitive at the scales demanded by neuroimaging, from large population cohorts such as the Human Connectome Project (HCP) to ex vivo ultra-high-field acquisitions at mesoscopic resolution. In this work, we present ExaTract, a parallelized spin-glass tractography framework designed for high-performance computing (HPC) architectures. Implemented as an open-source MPI+Kokkos application, ExaTract partitions the tractography domain into spatially overlapping regions processed across distributed memory nodes, with regular synchronization of boundary modifications to preserve global convergence guarantees. We validate ExaTract on both in vivo and ex vivo diffusion MRI data: the complete HCP dataset demonstrates population-scale throughput, while the high-resolution Chenonceau dataset provides the first mesoscopic structural connectome derived from global tractography.

\end{abstract}

\end{frontmatter}

\section{Introduction}

Tractography is a powerful technique for reconstructing the white matter pathways in the brain using diffusion MRI data. It has been widely used in neuroscience research and clinical applications to study brain connectivity and characterize various neurological disorders. Contrary to streamline tractography, which proceeds locally with the inference of axonal fibers independently, global tractography methods aim to solve the ill-posed problem of reconstructing the white matter connectivity from the local diffusion signals by considering the simultaneous inference of all streamlines \citep{poupon_regularization_2000,cointepas_spin_2002, kreher_gibbs_2008,reisert_global_2009}. As emphasized by \citet{mangin_2013}, this interest in global approaches has since become increasingly reflected with the large adoption of semi-global streamline filtering methods such as SIFT2 \citep{smith_sift2} and COMMIT \citep{daducci_commit}. Global tractography, rather than performing independent streamline reconstruction and filtering as separate steps, combines both within a single inference framework, yielding tractograms that are consistent with both the measured diffusion process and anatomical priors on a global scale.

Spin-glass tractography is a concrete implementation of global tractography \citep{poupon_regularization_2000}. Commonly based on a simulated annealing and Metropolis-Hastings algorithm, configurations of spin-glasses, corresponding to small portions of streamlines, are drawn from a proposal distribution and accepted or rejected based on a global energy function. The approach allows for the incorporation of global information and constraints, as the bending at white/gray matter (WM/GM) interfaces \citep{teillac} or the optimization of the diffusion model during spherical deconvolution \citep{kreher_gibbs_2008,reisert_mesoft_2014,konopleva_modelfree_2018}. In addition, the flexibility of the energy function in global tractography allows for the adaptation of the algorithm to any data modalities, making it a powerful tool for reconstructing brain connectivity at multiple scales \citep{axer_amunts}.

Despite these conceptual advantages, global tractography is computationally challenging. Its sequential nature creates memory and time bottlenecks that become prohibitive at the resolutions and dataset sizes encountered in practice. Large in vivo cohorts such as the Human Connectome Project (HCP) \citep{van_essen_Wu_2013} and the UK Biobank \citep{sudlow_ukb} provide millimeter resolution data across thousands of subjects, requiring efficient processing at the population level. At the other extreme, ex vivo ultra-high field diffusion MRI at the mesoscopic scale, and microscopy technique modalities, including 3D polarized light imaging (3D-PLI) \citep{axer}, polarization-sensitive optical coherence tomography \citep{wang_psoct}, and scattered light imaging \citep{menzel_sli}, enable the exploration of anatomical details of the WM pathways while disentangling the common biases of in vivo diffusion MRI, such as partial volume effects, gyral bias, and fiber crossings. The resulting Big Data, combined with the log-linear complexity of the global tractography algorithm, requires its implementation to scale up in order to tackle the aforementioned bottlenecks.

In this work, we introduce ExaTract, a parallelized implementation of the global spin-glass-based tractography technique designed to leverage high-performance computing (HPC) resources. ExaTract is implemented as an MPI+Kokkos application optimized for distributed memory architectures, enabling tractography at scales ranging from large population cohorts to a deep phenotyping approach at ultra-high resolution. The parallelization strategy relies on a spatial decomposition: the tractography domain is partitioned into sub-regions processed semi-independently across computing nodes, with overlapping ghost-cell regions and a communication protocol that synchronizes modifications across partition boundaries. This design is thought to preserve the global asymptotic consistency essential to the system convergence while enabling efficient parallel execution. We demonstrate the accuracy and the scalability of ExaTract on datasets representative of both in vivo population imaging and post-mortem exploration, covering the wide and deep phenotyping Big Data challenges.

\subsection{Related works}

A recent work from \citet{kruper} has proposed a survey of massively parallelized brain tractography frameworks pointing out the key information to be taken into account to design a suitable framework for HPC tractography. The main challenge in parallelizing the global tractography algorithm is the inherent sequential nature of its underlying Monte Carlo Markov chain. Further adaptations and assumptions were required to implement a parallelized architecture while maintaining acceptable efficiency. To date, a few models have been proposed to parallelize global tractography on shared-memory architectures. The parallelization opportunity comes from the observation that, among the possible modifications, the energy variations and the corresponding reconfigurations of spin-glasses remain local, even though the energy terms remain global.

\citet{philips_parallelized_2013} and \citet{wu_embarrassingly_2016} exploited this assumption by adopting an explicit parallelized global tractography scheme with two different approaches. \citet{philips_parallelized_2013} relied on a parcellation scheme to define disjoint, distant areas of authorized modification, each area handled by an independent thread. The distance to other areas prevents any corrupting interaction. At regular user-defined intervals, all threads stop their work and switch to the next area, introducing a cyclic scheme. \citet{wu_embarrassingly_2016} used a dynamic assignment of these areas. At regular intervals, the areas are dynamically recomputed. While these two techniques prevent both alteration of the cost function estimate and race conditions when two threads try to write a variable at the same time, they show limited scalability, as the effective area of processing for the global tractography is divided by more than the number of parallel workers.

\citet{konopleva_modelfree_2018} proposed an alternative strategy based on locking, which allows threads to safely access and modify the shared resource asynchronously. A computational challenge is the overhead introduced by mutex strategies, which increases significantly as the number of threads increases. In addition, it requires careful design to avoid deadlocks, in which two or more threads wait indefinitely for each other to release locks. This constraint practically limits the acceleration to a few threads at most.

To further leverage distributed-memory architectures, an adaptation of the global tractography algorithm is necessary. The very high dimensionality of the problem, comprising a few million spin-glasses (each with 6 variables), makes it a candidate for an asynchronous strategy based on spatial decomposition, where the tractography domain is shared between the computing units. While never tried in global tractography, the massive parallelization of simulated annealing has already been demonstrated for certain applications \citep{jayaraman_1988,greening_parallel}.

\section{Materials and Methods}

\subsection{Global tractography framework}

Spin-glass tractography simultaneously infers the entire tractogram through an optimization process constrained by the local orientation distributions to overcome the local perturbations. First introduced by \citet{poupon_regularization_2000} and further formalized by \citet{cointepas_spin_2002} and \citet{mangin_framework_2002}, particle-based global tractography involves particles, called spin-glasses, that represent small portions of streamlines. Each spin-glass extremity can connect with another spin-glass extremity to form chains of spin-glasses, ultimately leading to complete streamlines.

\subsubsection{System and energy definition}

A spin-glass $X_i$ is defined by its position $x_i$, its orientation $\mathbf{n_i}$, and its half-length $l$, here defined as the same for all the spin-glasses. Its extremities are designated by $X_i^{\alpha_i}$ where $\alpha_i = + 1$ for the one along $\mathbf{n_i}$ or $\alpha_i = - 1$ for the other along $-\mathbf{n_i}$. A generic extremity will then be denoted $X_i^\pm$. The entire system $\mathscr{M}$ is hence composed of

\begin{equation}
\mathscr{M} : \left\{
    \begin{array}{ll}
        \mathscr{X}: \Bigr\{ X_0 ... X_{N-1} \Bigr\} & \mbox{the spin-glasses,} \\ \\
        \mathscr{C}: \Bigr\{ (X_i^\pm, X_j^\pm) ... (X_k^\pm, X_l^\pm) \Bigr\} & \mbox{and their connections.}
    \end{array}
\right.
\end{equation}

In the following, $D$ denotes the data or the local orientation map, whichever local modeling is used. Historically, spin-glass-based global tractography relies on optimizing two energy terms, internal and external, such as the total energy of the system $E(\mathscr{M}, D)$ is given by:

\begin{equation}
\label{externalNoG}
    E(\mathscr{M}, D) = \alpha E_{int}(\mathscr{M}) + \beta E_{ext}(\mathscr{M}, D) \text{,}
\end{equation}

\noindent where $\alpha$ and $\beta$ represent an empirical weighting of the energies.

The \textbf{internal energy} tends to regularize connections between spin-glasses with respect to the expected axonal fiber anatomy, by optimizing the distance and incident angles between two connected spin-glasses. The energy term between two extremities, $U_{con}$ defined in \citet{reisert_global_2009} is commonly used as a more permissive and exploratory energy:
\begin{equation}
    \label{internalEnergy}
    U_{con}(X_1^{\alpha_1},X_2^{\alpha_2})=\frac{\left\| \mathbf{x_1} + \alpha_1 l \mathbf{n_1} - \mathbf{\Bar{x}} \right\|^2 + \left\| \mathbf{x_2} + \alpha_2 l \mathbf{n_2} - \mathbf{\Bar{x}} \right\|^2}{(l/2)^2} - L \text{,}
\end{equation}

\noindent where $\mathbf{\Bar{x}}$ is the midpoint between the two spin-glasses and $L$ is the connection likelihood. A large $L$ value strongly encourages the connection, while a small one promotes only connections between very close spin-glasses.

The total internal energy of the system is the sum of all the energy potentials corresponding to single connections:

\begin{equation}
    E_{int}(\mathscr{M}) = \sum_{(X_i^\pm, X_j^\pm) \in \mathscr{C}}{U_{con}(X_i^\pm,X_j^\pm)} \text{.}
\end{equation}

The \textbf{external energy} constraints the orientation of the spin glasses to align with the anatomical profile stemming from the measured data (dMRI, PLI, or ps-OCT...). In ExaTract, non-generative approaches were chosen, for which the alignment of any spin glass can be expressed as a function of the log-probability of its orientation stemming from the corresponding orientation distribution function (ODF), as defined in \citet{poupon_regularization_2000}:

\begin{equation}
    \label{externEnergies}
    E_{ext}(\mathscr{M}, D) = \sum_{X_i \in \mathscr{X}}{-\log \left( p(X_i) \right) } \text{,}
\end{equation}

\noindent where $p(X_i)$ is the probability to have an orientation $\mathbf{n_i}$ in the ODF at position $\mathbf{x_i}$. A further regularization term $\mu \mid \mathscr{X} \mid$ is added to the energy in order to control the number of spin-glasses and avoid any divergence of the optimization process.

\subsubsection{Optimization process}

Global tractography generally starts with a pseudo-random spin-glass configuration of the system. $n_{init}$ spin-glasses are randomly spread across the volume and oriented according to a Gibbs distribution sampler built upon the probabilities stemming from the local ODF, further sharpened or smoothed using a temperature $T_{gibbs}$. Through a reversible-jump Monte Carlo Markov chain (RJMCMC) process \citep{green_reversible_1995}, infinitesimal modifications are sequentially proposed to the system. Unlike traditional MCMC algorithms \citep{neal1993probabilistic}, RJMCMC allows the system to change its dimensions, meaning that a proposal can switch from one parameter space to another of a different dimension. This property makes RJMCMC particularly suitable for global tractography \citep{reisert_global_2009}, as the optimal number of particles needed to fit the data best is unknown in advance.

Propositions from the system configuration $\mathscr{M}$ to $\mathscr{M'}$ are drawn by randomly selecting a spin-glass and applying modifications, including:
\begin{itemize}
    \item \textbf{motion}: the spin-glass is randomly moved and reoriented, as in \citet{konopleva_modelfree_2018}. These random adjustments are selected from a normal distribution around the current position and orientation of the spin-glass, with a standard deviation $\sigma=0.05$. Low temperature allows local exploration of the configuration space, enabling the system to refine the alignment of spin-glasses with the local orientation distribution while allowing some degree of freedom to escape local minima.
    \item \textbf{connection}: all neighboring spin-glasses with a free extremity within a sphere of defined radius $r_{search}$ are considered for possible connection (see Figure~\ref{connection}). In addition, a no-connection possibility is also tested. This draws a distribution in which the connection proposal is chosen,

\begin{figure}[H]
\centering
\includegraphics[width=0.7\textwidth]{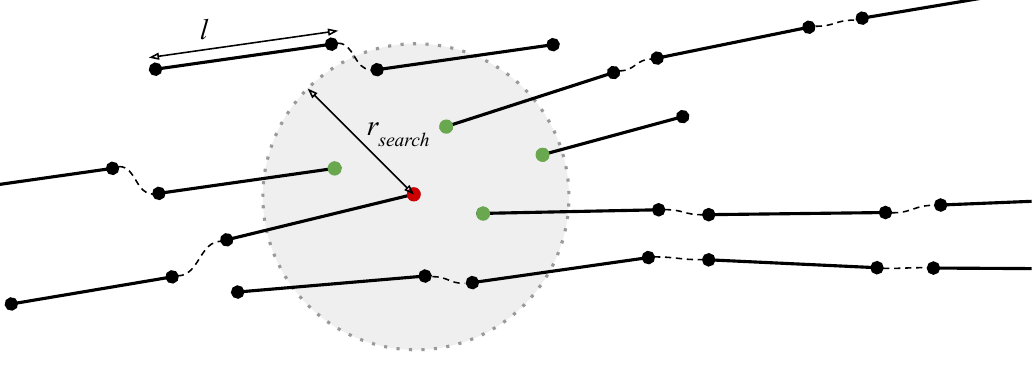}
\caption{Connection of a spin-glass. When proposing a connection, a random extremity of a random spin-glass, possibly already connected, is drawn. All possible connections with other spin-glass extremities found in a sphere of radius $r_{search}$ are tested. In addition, a no-connection possibility is also evaluated. The proposal is selected among the set of tested connections with a Gibbs sampler of temperature $T_{gibbs}$ based on their energy.}
\label{connection}
\end{figure}

    \item \textbf{creation}: a new spin-glass is randomly created within the tractography domain. This spin-glass orientation is chosen probabilistically using a Gibbs sampler similar to the one used for the initialization of the spin-glass map, ensuring that the proposed configuration is likely to be accepted while also maintaining enough randomness in the creation of new spin-glasses to explore a wide range of plausible configurations. To maximize the proposition likelihood, a connection is jointly proposed for the new spin-glass, which then has to be validated depending on the acceptance ratio $R$.
    \item \textbf{deletion}: the spin-glass is deleted, releasing its external energy and its connections.
\end{itemize}

If $p^{prop}(\mathscr{M'} | \mathscr{M} )$ designates the proposal distribution to switch from $\mathscr{M}$ to $\mathscr{M'}$, the modification of the configuration is proposed with an acceptance ratio of $R$, according to Bayes' rule and under the assumption of a Gibbs distribution of the system at temperature T:

\begin{equation}
    R = \frac{p^{prop}(\mathscr{M} | \mathscr{M'})}{p^{prop}(\mathscr{M'} | \mathscr{M} )} e^{- \frac{ \Delta E}{T} }\text{ .}
\end{equation}

The acceptance ratio only depends on the proposal distribution and the variation of energy $\Delta E$ between $\mathscr{M'}$ and $\mathscr{M}$. In the underlying Metropolis-Hastings optimization process, which is the sampling algorithm typically used to draw a Markov chain, a modification is accepted if $R \ge 1$. If $R < 1$, the modification is accepted with a probability of $R$.

This process is repeated a large number of times until a stopping criterion is met, following a temperature-cooling schedule. The performed modifications tend to minimize the system energy $E(\mathscr{M}, D)$, leading the system configuration to converge towards an optimal tractogram that is assumed to provide a plausible configuration both from the point of view of the anatomy and the data. The final tractogram is then extracted from the optimal configuration of spin glasses by following the connections between them.

\subsection{Parallelization strategy of ExaTract}

Each modification brought to a spin-glass $X_i$, initially connected to spin-glasses $X_j$ and $X_k$ and potentially newly connected to spin-glasses $X_{j'}$ and $X_{k'}$, can be rewritten locally under the following general form:

\begin{equation}
\begin{split}
    \Delta E & = U_{con}({X'}_{i}^{+},X_{j'}^{\pm}) - U_{con}(X_i^{+},X_{j}^{\pm}) \\
    & + U_{con}({X'}_{i}^{-},{X}_{k'}^{\pm}) - U_{con}(X_i^{-},X_{k}^{\pm}) \\
    & - \log \left( p({X'}_{i}) \right) + \log \left( p(X_i ) \right) \\
    & + \gamma \mu \text{ ,}
\end{split}
\end{equation}

\noindent where $\gamma$ is 0 in case of motion or connection modification, 1 in case of creation, and -1 in case of deletion. This local formulation of the energy variation is a key property that enables parallelization of the global tractography algorithm.

\subsubsection{Domain decomposition}

To fit with current parallel architectures, the tractography domain is divided into $N_{parcel}$ precomputed static sub-domains using the METIS library \citep{karypis_metis}. The binary mask corresponding to the tractography domain can be represented as a graph, in which each voxel corresponds to a vertex and each edge represents a neighbor relationship between two adjacent voxels. The multilevel k-way partitioning algorithm \citep{karypis_metis_1997} provided by the METIS toolbox decomposes this graph, taking into account the spatial locality of the data. The partitioning process aims to create sub-domains that are as independent as possible by following the morphology of the tractography domain while minimizing the number of edge cuts between sub-domains. Each partition being assigned to a specific shared-memory computing unit, the computed domain decomposition directly contributes to minimizing communication costs between nodes during spin-glass optimization. A strong constraint in the partitioning process is to ensure that the sub-domains are sufficiently large to allow for efficient parallel processing while also being small enough to fit within the memory of each node, which further adds to the complexity of the parallelization of global tractography.

\begin{figure}[H]
\centering%% For center alignment of image.
\includegraphics[width=1.0\columnwidth]{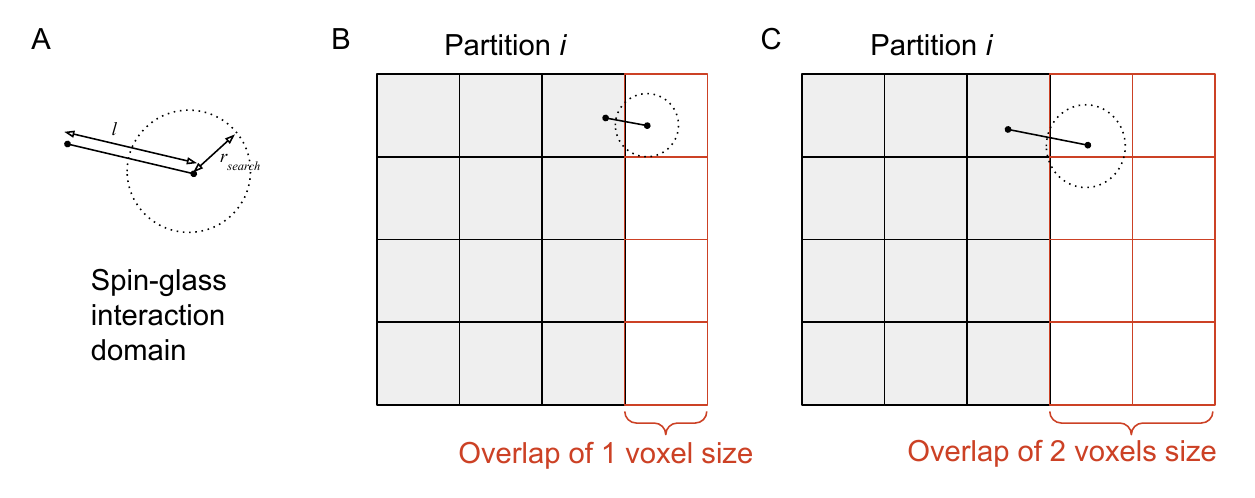}
\caption{2-D examples of overlap depending on $l$, $r_{search}$ (A) and the voxel size. When $l + r_{search}$ is lower than the voxel size (B), the overlap pattern includes the 26 directly adjacent voxels. When $l + r_{search}$ is greater than the voxel size (C), the external overlap of the partition $i$ needs to be at least two voxels wide. The same logic applies symmetrically on the other side of the boundary, so the minimum size of the full overlap is 2 voxels.} 
\label{overlap}
\end{figure}

To ensure system continuity across partition boundaries, an overlap must be defined between adjacent sub-domains. This overlap allows information exchange between nodes and ensures that modifications made in one sub-domain can influence neighboring sub-domains, which is crucial to preserve the continuity of the underlying Markov field. A set of ghost cells \citep{kjolstad_ghost_2010} has to be defined within each overlapping area and has to be updated at regular intervals to reflect the changes made in the neighboring sub-domains. The width of the overlap is determined by the spin-glass length and the maximum interaction radius of spin-glasses, ensuring that all relevant interactions are captured while minimizing communication overhead (see Figure \ref{overlap}).

Gathering modifications in the overlap areas at regular intervals is a necessary assumption for adapting the underlying RJMCMC process to a shared memory environment, based on the observation that a single modification affects only a local area of the system, even if the energy terms are global. By regularly updating ghost cells at a certain frequency, we can ensure that modifications made in one sub-domain are reflected in neighboring sub-domains, ensuring coherent convergence across overlap areas and, consequently, across the entire domain.

The frequency of communication across the node is then defined using an additional methodological parameter, $\omega$, representing the workload per iteration, and corresponding to the number of proposals performed before synchronizing the ghost cells. The workload $\omega$ is user-defined in order to adapt to the size of the subdomain and ensure a balanced advancement of the system across the entire domain. $\omega$ is directly related to the communication frequency, as a higher workload corresponds to a lower communication frequency. This approach allows us to maintain the system's global coordination while enabling efficient parallel processing of the tractography algorithm.

\subsubsection{Multithreading}

In addition to domain decomposition, each sub-domain assigned to a node can leverage the node's multithreading capabilities to perform concurrent execution of the tractography algorithm within that sub-domain. While previous implementations based their parallelization on scheduling and synchronization of the modifications \citep{philips_parallelized_2013,wu_embarrassingly_2016}, or mutual exclusion methods \citep{konopleva_modelfree_2018}, we introduce an additional, but less restrictive assumption to allow threads to make changes in parallel without incurring any additional overhead. Threads independently draw modifications and apply them concurrently to the spin-glass maps. This assumption is expected to be less stringent than the previous one done in domain decomposition.

In computer science, modifying the same memory location at the same time is called a race condition. It may lead to undefined behavior, as the memory location may not hold the intended value. In ExaTract, we avoid the potential fatal consequences of race conditions by a posteriori corrections: before communications, the spin-glass map configurations are checked to correct for abnormalities. While an additional linear-complexity step is required to implement it, the added value of thread asynchronism is relevant to minimize computing time.

\begin{figure}[H]
\centering%% For center alignment of image.
\includegraphics[width=0.65\columnwidth]{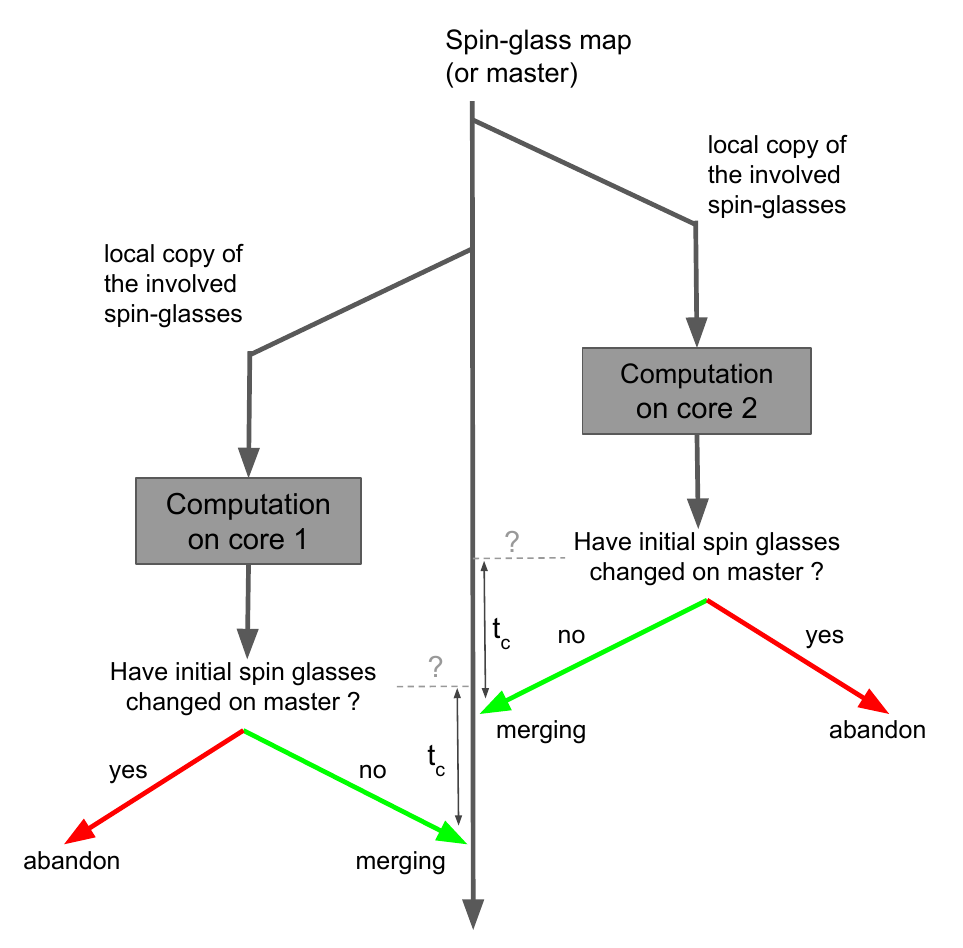}
%% Use \caption command for figure caption and label.
\caption{Local copy system for parallel computing of modifications. This system includes (a) a minimal copy of the involved spin-glass, (b) parallel computation on the different threads, and (c) a quick check of whether the original spin-glass has changed. While allowing for parallel processing, a concurrent situation can only occur during the time lapse $t_c$.}\label{figLocalCopy}
\end{figure}

A local and reduced copy of the system $\mathscr{M}$ is developed in ExaTract to mitigate race condition occurrence. The different modification proposals are evaluated on parallel branches, each operating on a private and temporary copy of the current spin-glass configuration around the spin-glass being modified. As illustrated in Figure \ref{figLocalCopy}, this approach minimizes the duration of concurrent access $t_c$ to shared data while fully preserving parallel execution. A model for the distribution of conflict occurrences is presented in \ref{app1}.

\subsubsection{Conflict resolution}
\label{conflictResolution}

The same correction mechanism is applied in ghost cells, where similar conflicts can occur when merging modifications stemming from two adjacent nodes. The conflict resolution follows a deterministic, symmetrical strategy to ensure the same correction is applied to all nodes sharing the ghost cell (for more details, see \ref{app2}).

This approach enables rapid recovery from conflicts by efficiently restoring the system to a consistent state. Taking these anomalies into account at runtime allows us to preserve the asymptotic convergence towards an optimal configuration of the spin-glasses. However, conflicts should be minimized as much as possible, as they break the Markovian framework. This minimization is possible by reducing $\omega$, the number of operations performed per iteration, even though it also plays in favor of an increase in communication events. It raises the question of the optimal balance between communication and conflict occurrence, which is crucial for the computational efficiency of the parallelization strategy but also for the convergence of the system, as conflicts are expected to slow down the convergence.

\begin{figure}[H]
\centering%% For center alignment of image.
\includegraphics[width=1.0\textwidth]{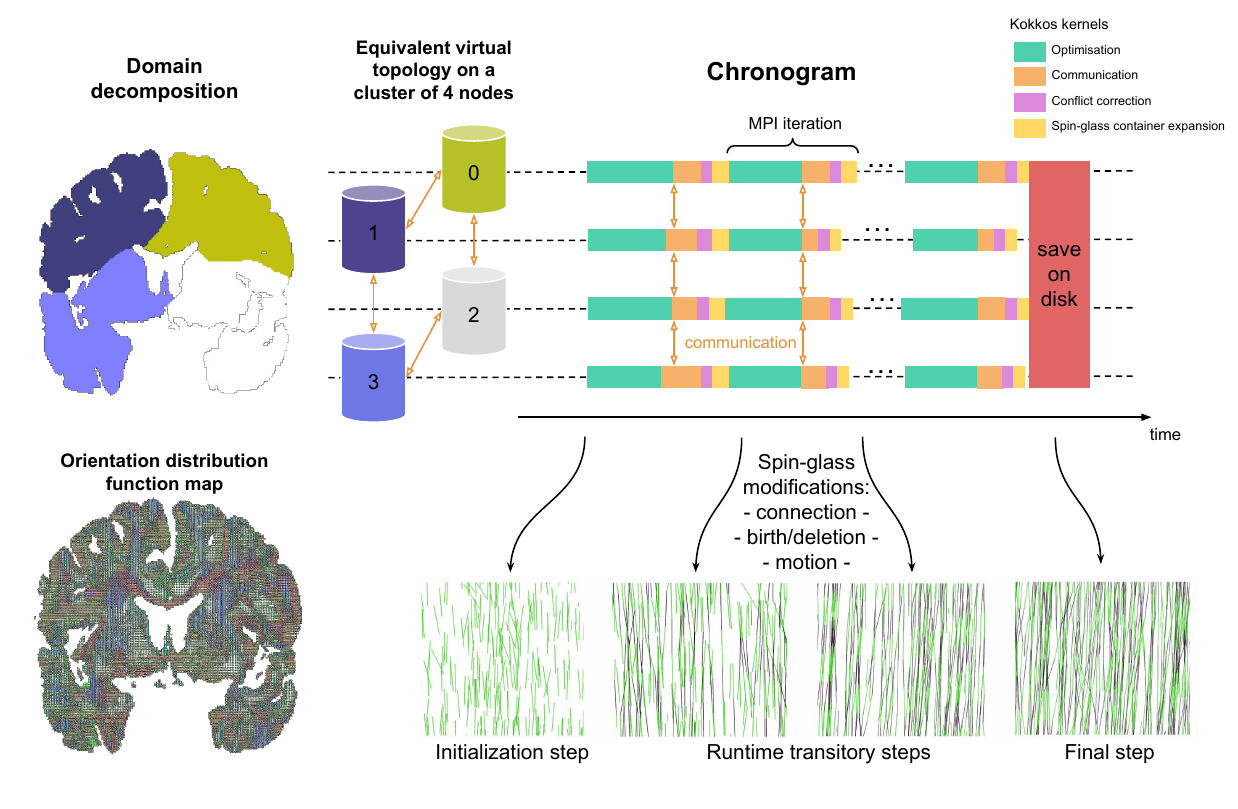}
\caption{Chronogram of ExaTract execution. A brain, along with its ODF map, is divided into four partitions, each assigned to a node of a supercomputer. The four MPI processes run global tractography in a synchronized way, performing a user-defined number of operations defined by $\omega$. An MPI iteration comprises (1) an optimization kernel that effectively applies the global tractography on its partition, (2) communication steps that exchange the modifications that occurred in overlapping areas, (3) a conflict correction step, and (4) the spin-glass map expansion.}\label{chronogram}
\end{figure}

\subsection{Implementation and monitoring}

ExaTract is fully implemented in the C++ language using object-oriented programming. It employs data structures, IO readers and writers, and dMRI processing objects, including diffusion ODFs and subsequent methods, as part of the Ginkgo C++ toolkit\footnote{https://framagit.org/cpoupon/gkg}.

MPI communication is managed with \texttt{Isend} and \texttt{Irecv}, which are asynchronous, non-blocking methods that allow communication to overlap with other computation. Multithreading is implemented using Kokkos~\citep{carter_2014}, an open-source portability library. Kokkos abstracts parallel execution through two core concepts: \textit{kernels}, which encapsulate parallel loops, and \textit{views}, which manage memory allocations across architectures.

ExaTract required several Kokkos kernels to execute its parallel workload: spin-glass map initialization, global tractography optimization, a posteriori correction, merging communicated modifications, and a set of kernels for dynamic expansion of the spin-glass map. Because the total number of spin-glass segments per voxel is not known at the beginning, ExaTract supports dynamic allocation of new spin-glasses within the Kokkos view representing the spin-glass map. It is designed to favor cache efficiency when iterating over neighboring spin-glasses, which is especially necessary for high-resolution datasets. Algorithm~\ref{algo:exatract} and Figure~\ref{chronogram} summarize the chronological execution of these kernels.

ExaTract comes with additional features, including a logging system for real-time monitoring of algorithm convergence and error occurrence analysis. Regular save checkpoints are done as ExaTract might need several consecutive jobs to reach convergence when targeting high-resolution datasets. In addition, the backup ensures resilience in case of an unexpected job failure.

\begin{algorithm}
\caption{Simplified pseudocode of the ExaTract process}
\label{algo:exatract}
\begin{algorithmic}[1]
\For{$it = 0 \to it_{max}$} \Comment per MPI process
    \For{$draw = 0 \to \omega*N_{voxel} / N_{threads}$} \Comment per thread
        \State draw a random $spin\text{-}glass$
        \State draw a modification proposal
        \If{ $p \le R$ }
            \State apply modification
        \EndIf
    \EndFor
    \For{$spin\text{-}glass$ in local map} \Comment parallelized over threads
        \If{ $spin\text{-}glass$ is abnormal }
            \State correct $spin\text{-}glass$
        \EndIf
    \EndFor
    \For{$spin\text{-}glass$ in local overlap}
        \If{ $spin\text{-}glass$ has changed }
            \State add $spin\text{-}glass$ to $message$
        \EndIf
    \EndFor
    \For{ all neighbor ranks $i$} \Comment parallelized over threads
        \State send $message$
        \State receive $message_{received}$
    \EndFor
    \For{$spin\text{-}glass$ in $message_{received}$}
        \State apply modification $spin\text{-}glass$
    \EndFor
    \For{$voxel$ in local map} \Comment parallelized over threads
        \State expansion of $voxel$ if necessary
    \EndFor
    \State $T = T + \delta T $ 
\EndFor
% \State \Return $array\_slides$
\end{algorithmic}
\end{algorithm}

\subsection{Experimental protocols}

To assess the consistency and the scalability, ExaTract was tested on both synthetic and real-case datasets of varying sizes (see Table \ref{tab:datasets}). ExaTract has been designed to be compatible with any kind of local model. In the frame of this application, we used the analytical Q-ball model \citep{descoteaux_fast_2007} computed with a Laplace-Beltrami regularization factor of 0.006. All ODFs were sampled using 10,000 uniformly distributed orientations over the unit sphere, corresponding to a solid angular resolution of 2.0 degrees. For all experiments, creation, deletion, motion, and connection probabilities ratios were set to 0.25, 0.02, 0.25, and 0.15, respectively. The connection likelihood $L$ was set to 0.5, and the regularization of spin-glass count $\mu$ was set to 0.1. The internal and external energy weighting $\alpha/\beta$ was set to 5.0/4.0. The spin-glass length was typically set slightly below the resolution in order to minimize overlap size. These parameters were either derived from the literature \citep{reisert_global_2009,konopleva_modelfree_2018} or empirically defined based on preliminary experiments to ensure convergence of the system while allowing for exploration of the configuration space.

\subsubsection{Synthetic data}

The DiSCo MRI phantom \citep{rafael-patino2021diffusion} was used as a synthetic dataset to evaluate the performance of ExaTract with different numbers of threads and MPI processes on the same dataset. Simulated data of a shell with a b-value of 3000 $s.mm^{-2}$ was used, and the tractography domain was defined as the whole phantom. The following parameters were used for ExaTract: 10 initial spin-glasses per voxel, spin-glass length and connection radius of $l=r_{search}=0.99$ mm, 500 MPI iterations with $\omega = 10$ operations per voxel and per MPI iteration, an exponential decreasing cooling schedule from $T_{init} = 0.4K$ to $T_{final} = 0.03 K$.

\subsubsection{In vivo cohort}

The 1065 subjects of the HCP database \citep{van_essen_Wu_2013} were processed with ExaTract using the following parameters: 10 initial spin-glasses per voxel, spin-glass length and connection radius of $l=r_{search}=1.249$ mm, 14000 MPI iterations with $\omega = 1$ operation per voxel and per MPI iteration, exponential decreasing cooling schedule from $T_{init} = 0.4K$ to $T_{final} = 0.01 K$. The preprocessing comprised the application of the FreeSurfer\footnote{https://surfer.nmr.mgh.harvard.edu/} pipeline to compute masks and the Ginkgo pipeline\footnote{https://framagit.org/cpoupon/gkg/} to compute a map of ODFs as described in \citet{herlin_sex-related_2024}.

\subsubsection{Post-mortem high-resolution datasets}

The Chenonceau dataset \citep{legeay} consists of an entire cerebrum scanned with a preclinical MRI system at 11.7T (Bruker Biospin, Ettlingen), at an isotropic resolution of 200 µm according to a multiple shell sampling of the q-space including 25 directions at $b=1000s.mm^{-2}$, 60 at $b=4500s.mm^{-2}$ and 90 at $b=8000s.mm^{-2}$. The tractography domain was defined as the entire brain; the domain was split into 800 parcels of $144414 \pm 7$ voxels. The ODF map was computed using the 90 directions of the highest b-value shell. An additional dataset of a macaque brain (Macaca Fascicularis species) scanned at 250 µm isotropic resolution with the same diffusion MRI protocol \citep{macaque} was also used to evaluate the scalability of ExaTract.

\begin{table}[H]
\centering
\begin{tabular}{l | r r r r}
\hline
  Dataset & DiSCo & HCP subject & Macaque & Chenonceau \\
  \hline
  Type & phantom & in vivo human & ex vivo specimen & post-mortem human specimen \\
  Isotropic resolution & 1 mm & 1.25 mm & 250 µm & 200 µm \\ 
  Dataset size (voxels) & 15040 & $5-8\text{x}10^5$ & $3.26\text{x}10^6$ & $1.15\text{x}10^8$ \\
  \hline
\end{tabular}
\caption{Sizes of the different use-case datasets.}
\label{tab:datasets}
\end{table}

\subsubsection{Hardware}

For benchmarking on the DiSCo dataset, ExaTract was run using a desktop machine equipped with 8 bi-threaded CPU cores (2.40GHz, cache 16Mo). For all other datasets, the Irène-Rome partition of the Joliot-Curie supercomputer of the TGCC (GENCI, Très Grand Centre de Calcul, CEA, Bruyères-le-Châtel, France) was used, mobilizing up to 200 Non-Uniform Memory Access (NUMA) nodes of 64 bi-process cores (256 Go RAM per node, 2.60GHz) to compute the global tractography. Due to the non-uniform memory access in this architecture, the number of MPI processes per node was typically set to 8, favoring local memory of the multi-threaded processes, as further demonstrated in section~\ref{labelling}.

\section{Results}

\subsection{Parallelization assumption validation}

The design of ExaTract relies on two main assumptions to allow for efficient parallelization: (1) the modifications performed in a voxel only impact a local area of the system so that domain decomposition is effective, and (2) the occurrence of conflicts in multithreading and in overlapping regions remains negligible in the convergence of the system. To validate these assumptions, we first estimated the mean conflict occurrence in both cases and then compared it with the observed occurrences across the different benchmarks.

\begin{figure}[H]
\centering
\includegraphics[width=0.7\columnwidth]{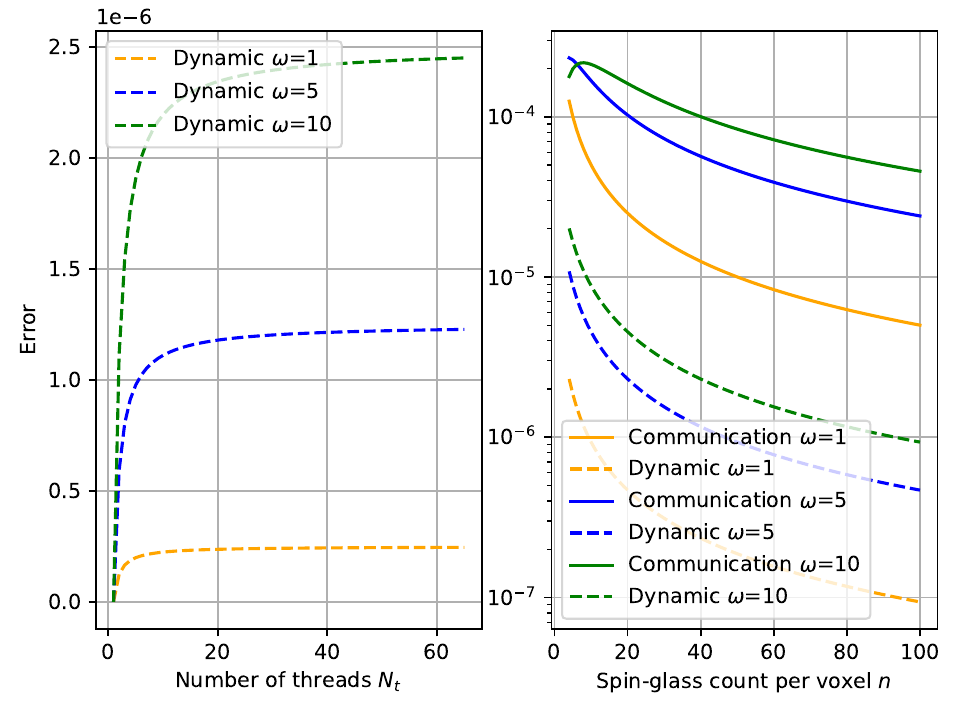}
\caption{Estimation of error occurrences per voxel evaluated over multiple $\omega$ values. On the left, the relationship between the multithreading conflict occurrences and the number of threads $N_t$ shows the convergence of equation \ref{communication_conflict} to an upper bound weighted by $\omega$ at high thread counts. On the right, dependencies of the two conflict sources on the number of spin-glasses per voxel show the preponderance of communication conflicts over multithreading conflicts (or dynamic conflicts) ($N_t$ was fixed to 16). For the numerical applications, $\overline{R}$ was fixed to 0.1, $t_c$ to 0.001 and $p_{overlap}$ to 0.05. Communication conflicts occurring are counted per voxel in overlapping areas.}
\label{figError}
\end{figure}

Based on equations \ref{multithreading_conflict} and \ref{communication_conflict}, we estimated the upper limit of error occurrences from the two parallelization layers (Figure \ref{figError}). The estimation of communication conflict occurrences is based on the probability that a proposed modification will affect another modified spin-glass in the overlapping area of another node. The estimation of multithreading conflicts is based on the probability that two threads draw and modify the same spin-glass, which is directly related to the number of threads and the number of operations per MPI iteration. Both estimations are weighted by the squared mean acceptance ratio $\overline{R}^2$, since a conflict can occur only if both proposed modifications are accepted.

Numerical applications of the equations established in \ref{app1}, exposed in Figure \ref{figError}, show that the errors stemming from domain decomposition are more likely to occur than the ones stemming from multithreading. A striking result is that multithreading conflicts converge to an upper bound as the number of threads increases. The latter comes from an equilibrium between the threads added and the distribution of the workload. The distribution of work across threads tends to offset the increase in the number of thread pairs exactly and, consequently, the number of conflicts.

% This estimation is consistent with the observed error occurrences in the following benchmarks, where the number of multithreading conflicts is negligible compared to the communication conflicts.

\subsubsection{Validation on the DiSCo phantom}

To assess the effects of ExaTract parallelization, we run ExaTract on the DiSCo phantom using four parallelization strategies and compare with a baseline computed without parallelization. The effects of multithreading were evaluated on the non-partitioned domain using 16 threads in parallel. The effects of an extreme domain decomposition were isolated using a 16-partitioned domain (1 thread per partition) with $\omega$ = 10 op/voxel. Two intermediary experiments were conducted using 2 and 4 partitions, with 8 and 4 threads per partition, respectively. Global tractography is a stochastic optimization such that a single run differs from another when initialized with different seeds. To compare the variance due to the algorithm's own randomness with the effects of parallelization, we computed 10 seed-independent runs per parallelization configuration.

\begin{figure}[H]
\centering
\includegraphics[width=1.0\columnwidth]{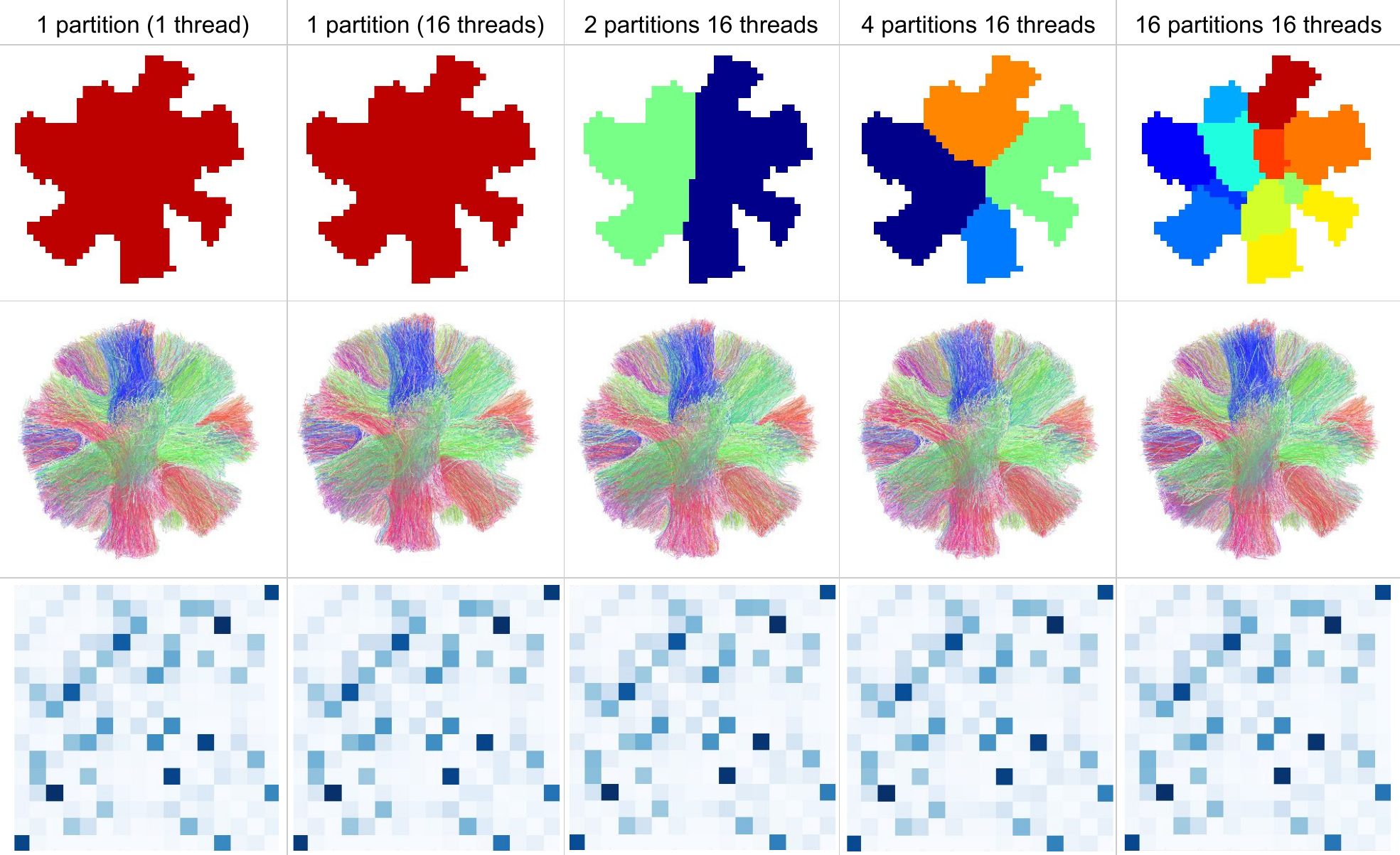}
\caption{Effects of the parallelization strategies on the DiSCo phantom using different domain decompositions and multithreading approaches while using a single seed per computation. Regardless of the decomposition scheme (top) and the computational resources allocated for multithreading, the resulting tractograms (middle) and their corresponding structural connectivity matrices (bottom) remain consistent.}\label{discoMAchine}
\end{figure}

The obtained tractograms, compared qualitatively in Figure \ref{discoMAchine} and quantitatively in Table \ref{DiSCobenchmarkTable} in terms of fiber and spin-glass counts, show very similar fiber configurations, even for the extreme case of domain decomposition.

In Table \ref{DiSCobenchmarkTable}, we additionally report conflict occurrences. Conflict occurrences are consistent with previous estimates, with less than 0.07\% of total operations in conflict in the extreme case of domain decomposition. As previously modeled, the multithreading conflicts are negligible compared to the communication conflicts, with a maximum of $9.73\text{x}10^{-5}\%$ of the operations being in conflict for the extreme case of multithreading. We note that conflict occurrences vary from one experiment to another while using 16 threads in total (independently of the partitioning). It might come from the extreme cases of the parallelization on a very small dataset, where 1) overlapping areas are likely to be distributed over more than 2 MPI processes, which increases the probability of communication conflicts, and 2) the reduced dataset size leads to a higher probability for two threads to be synchronized (so that assumptions made in \ref{app1} are not valid) and draw to the same spin-glass, which increases the probability of multithreading conflicts. Despite these discrepancies, the observed conflicts remain negligible and do not significantly affect the system's convergence.

Since the DiSCo phantom is a synthetic dataset, the obtained tractograms were also compared to the ground truth using ROC AUC, which evaluates both tractogram sensitivity and accuracy \citep{girard_2023}. The ROC AUC stability indicates a strong reproducibility in the inferred tractograms across parallelization strategies. With a maximum standard deviation of 0.005, the ROC AUC distributions per experiment illustrate the robustness of the parallelization strategy to find consistent results from different initializations.

To further compare the seed-induced variability to the parallelization-induced variability, we computed the Pearson correlations $r_{seed\; to\; seed}$ between the connectivity matrices obtained with different seeds in the same parallelization configuration and the Pearson correlations $r_{parallel\; to\; sequential}$ between the connectivity matrices obtained in a given parallelization strategy to the baseline connectivity matrices computed sequentially. 

$r_{seed\; to\; seed}$ is very high, with a minimum value of 0.9977. This indicates that the variability induced by the stochastic nature of the algorithm is low and that the obtained tractograms are consistent across different runs. The comparison of the tractograms obtained with different parallelization strategies to the baseline tractogram computed without parallelization also shows a very high Pearson correlation $r_{parallel\; to\; sequential}$, with a minimum value of 0.9978. This indicates that the parallelization strategies do not significantly affect the resulting tractograms and that the obtained results are consistent across different parallelization configurations.

To assess whether partioning and multithreading parallelizations alter the connectivity beyond the inherent random floor of the global tractography optimization, we compared pairwise connectivity matrix correlations within the experiments of the same implementation (different seeds) and pairwise connectivity matrix correlations between the sequential and the parallel implementations. Because pairwise comparisons drawn from a shared pool of runs do not follow the standard assumptions of independent samples, we used a label-permutation test (10,000 resamples), which found no evidence that parallelization introduces divergence beyond the algorithm's own noise: as reported in Table \ref{DiSCobenchmarkTable}, all p-values were found greater than 0.05, rejecting the null hypothesis that the two distributions are different.

\begin{table}[H]
\centering
\begin{tabular}{l | r r r r r r}
    \hline
  Partitioning & 1 & 1 & 2 & 4 & 16 \\
  Threads per partition & 1 & 16 & 8 & 4 & 1 \\
  Overlap coverage ($\%$) & 0 & 0 & 7 & 18 & 47 \\
  \hline
  \hline
%   Fiber count      & 26538  & 26784  & 26801  & 26965  & 26862 \\
  Fiber count      & $26662$  & $26653$  & $26789$ & $26792$  & $26906$ \\
   & $\pm 87$  & $\pm 73$  & $\pm 78$ & $\pm 105$  & $\pm 77$ \\
\hline
  Spin-glass count & 548875 & 548470 & 550974 & 551556 & 551517 \\
                   & $\pm$  1223 & $\pm$  1450 & $\pm$  1089 & $\pm$  1582 & $\pm$  899 \\
\hline
%   Conflicts      & 0 & 0 & 2041 & 6563 & 21962 \\
%   Multithreading Conflicts      & 0 & 4 & 32 & 86 & 211 \\
%   Multithreading conflict occurence ($\text{x}10^{-7} \%$)      & 0.0 & 0.53 & $4.25$ & $11.4$ & 0 \\
  Multithreading conflicts       & 0.0 & 0.39 & 4.49 & 9.73 & 0.0 \\
  ($\text{x}10^{-7}$)       & 0.0 & $\pm$ 0.17 & $\pm$ 0.67 & $\pm$ 1.07 & $\pm$ 0.0 \\
%   Communication Conflicts      & 0 & 0 & 2009 & 6477 & 21751 \\
%   Communication conflict occurence ($\text{x}10^{-4} \%$)      & 0.0 & 0 & $3.81$ & $4.79$ & $6.15$ \\
\hline
  Communication conflicts      & 0.0 & 0.0 & 3.79 & 4.70 & 6.13 \\
   ($\text{x}10^{-4}$)     & $\pm$ 0.0 & $\pm$ 0.0 & $\pm$ 0.14 & $\pm$ 0.09 & $\pm$ 0.05 \\
%   Computation time (s)& 5283.7 & 725.62 & 740.9 & 751.35 & 877.6 \\
%   VC/(VC+IC) & 0.710 & 0.716 & 0.709 & 0.709 & 0.709 \\
%   ROC AUC & 0.918 & 0.922 & 0.924 & 0.916 & 0.923 \\
\hline
  ROC AUC & 0.922 & 0.923 & 0.921 & 0.922 & 0.922 \\
          & $\pm$ 0.003 & $\pm$ 0.003 & $\pm$ 0.005 & $\pm$ 0.004 & $\pm$ 0.005 \\
%   Pearson correlation & 1.0 & 0.998 & 0.998 & 0.998 & 0.998 \\
%   Accuracy (at 1\% threshold) & 0.949 & 0.949 & 0.949 & 0.949 & 0.949 \\
  \hline
  $r_{seed\; to\; seed}$ & 0.9977 & 0.9980 & 0.9982 & 0.9979 & 0.9980 \\
  (45 pairs)                                  & $\pm 0.0005$ & $\pm 0.0004$ & $\pm 0.0003$ & $\pm 0.0005$ & $\pm 0.0004$ \\
  \hline
  $r_{parallel\; to\; sequential}$ & - & 0.9978 & 0.9979 & 0.9978 & 0.9979 \\
  (100 pairs) & - & $\pm 0.0004$ & $\pm 0.0004$ & $\pm 0.0005$ & $\pm 0.0004$ \\
  \hline
  Permutation test p-value & - & 0.331 & 0.408 & 0.147 & 0.940 \\
  \hline
\end{tabular}
\caption{Statistics of the ExaTract benchmark on the DiSCo dataset. To assess reproducibility, all experiments were repeated 10 times. Multithreading conflicts are reported as the ratio of corrupted operations out of the total operations performed. Communication conflicts are reported as the ratio of corrupted operations among total operations performed in overlapping regions.}\label{DiSCobenchmarkTable}
\end{table}

\subsubsection{Validation on an in vivo HCP human subject}
\label{labelling}

\begin{figure}[H]
\centering
\includegraphics[width=1.0\columnwidth]{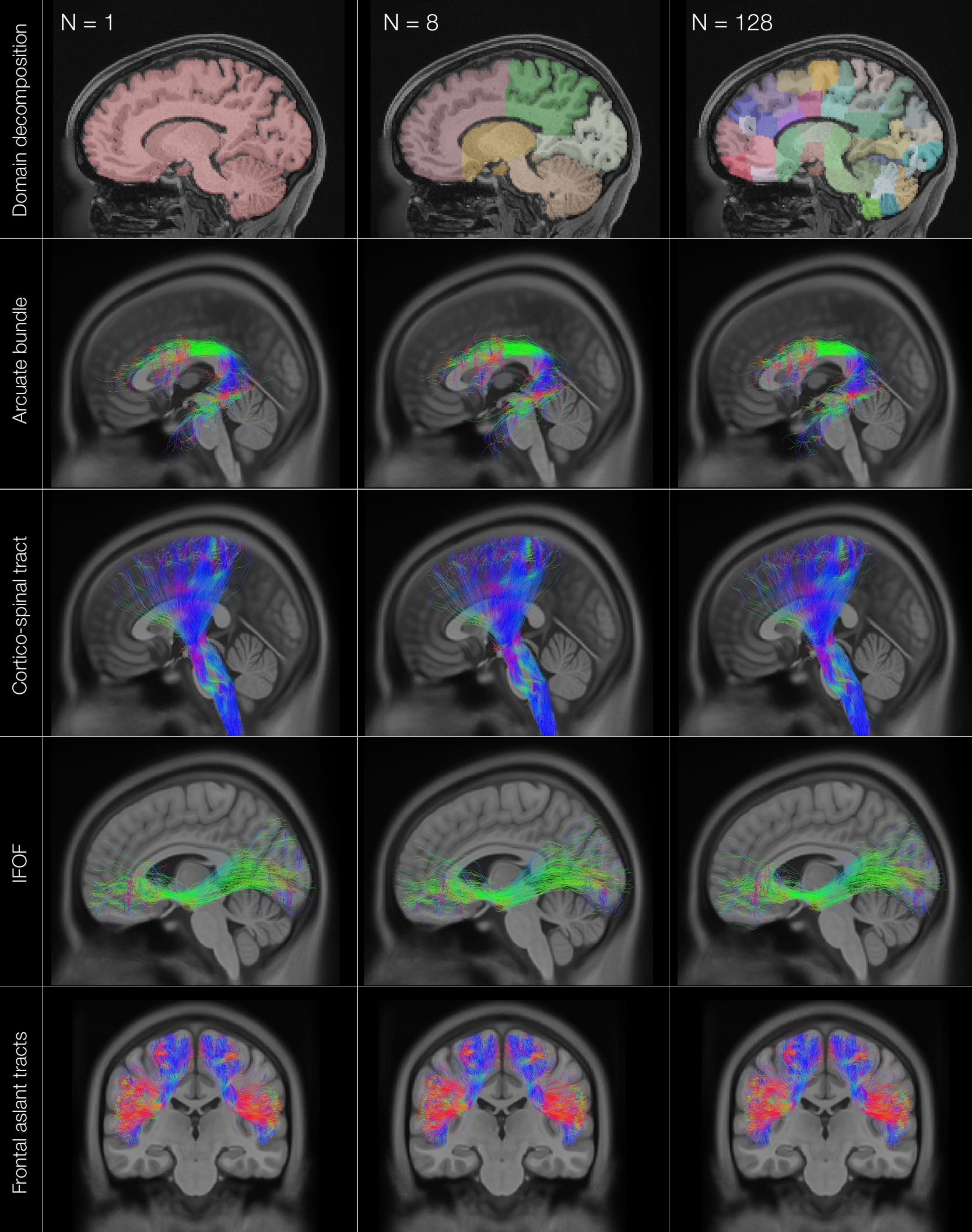}
\caption{Fiber labeling results on one subject of the HCP database from three ExaTract runs computed with 32 ($N=1$ partition), 256 ($N=8$ partitions), and 4096 ($N=128$ partitions) threads, on the arcuate bundle, the corticospinal tracts, the inferior fronto-occipital fasciculus, and the frontal aslant tracts. The shape of the recovered bundles is similar from one run to another.}
\label{HCPbundle}
\end{figure}

Similarly, five ExaTract runs were launched on one subject of the HCP database with 1, 8, 32, 64, and 128 partitions, multithreaded over 16 bi-threaded cores per partition. The obtained tractograms were automatically labeled, resampled in the MNI space, to extract the deep WM bundles based on the deep WM bundle atlas established in \citet{herlin_sex-related_2024}. As compared in Figure~\ref{HCPbundle}, the shapes of the labeled bundles are similar from one run to another, even in the extreme case of 128 partitions. The obtained tractograms were also compared in terms of fiber and spin-glass count, and conflict occurrences (Table \ref{HCPbenchmarkTable}). The number of fibers and spin-glasses remains stable with the number of partitions, which is consistent. The occurrence of communication conflicts is consistent with the previous estimate, and is almost constant over the experiment, with a maximum of 0.023\% of the operations in conflict for the extreme case of domain decomposition. Despite these conflicts, the obtained bundles are qualitatively similar, suggesting that the system's convergence is not significantly affected.

Regarding computation time, a significant acceleration is observed when increasing the number of partitions, with a speed-up of 88.4 using 128 partitions compared to the single-partition case. The parallelization efficiency remains high, at 0.69 for 128 partitions with overlapping regions covering 47\% of the domain, indicating that the communication and synchronization overhead remains acceptable for the considered workload.

\begin{table}[H]
\centering
\begin{tabular}{l | r r r r r r}
  \hline
  Partitioning & 1 & 8 & 32 & 64 & 128 \\
  \hline
  Overlap coverage ($\%$) & 0 & 9.2 & 21 & 31.88 & 45.1 \\
  Fiber count      & 886186  & 889966  & 891224  & 894942  & 899740 \\
%   Spin-glass count & 29416185 & 29610744 & 29715095 & 29789227 & 29968439 \\
  Spin-glass count & $2.94\text{x}10^7$ & $2.96\text{x}10^7$ & $2.97\text{x}10^7$ & $2.98\text{x}10^7$ & $3.0\text{x}10^7$ \\

    % Dynamic Conflicts            & 2 & 1517 & 3842 & 5518 & 7754 \\
    Multithreading conflicts  ($\text{x}10^{-7}$)          & $2.35\text{x}10^{-3}$ & 1.78 & 4.51 & 6.48 & 9.11 \\
%   Communication Conflicts      & 0 & 140205 & 372089 & 566373 & 870247 \\
    Communication conflicts  ($\text{x}10^{-4}$)     & 0 & 1.79 & 2.08 & 2.09 & 2.27 \\
%   Computation time (s)& 1044 & 159.59 & 44.59 & 22.91 & 11.81 & 877.6 \\
  Computation time& 17h24 & 2h40 & 44min35s & 22min54s & 11min48s \\
  Speed-up & 1 & 6.5 & 23.4 & 44.6 & 88.4 \\
  Efficiency & 1 & 0.82 & 0.73 & 0.71 & 0.69 \\
    \hline
\end{tabular}
%% Use \caption command for table caption and label.
\caption{Tractograms statistics, conflict occurrences, and performance of ExaTract benchmarked on one subject of the HCP dataset. 16 bi-threaded cores are used for each partition.}\label{HCPbenchmarkTable}
\end{table}

\subsection{Scalability benchmarking at ultra high resolution}
\label{benchmarkParagraph}

The scalability of ExaTract was evaluated under two perspectives: strong and weak scaling. In the first case, scalability is evaluated by measuring parallelization efficiency on a fixed-size problem, while in the second, it is evaluated on a growing dataset while preserving the workload per processor.

To evaluate the strong scaling, ExaTract was run on the ex vivo macaque brain dMRI dataset using increasing numbers of cores while keeping the entire tractography domain fixed. For each combination, the speed-up of the first 20 ExaTract iterations was measured for three different workloads ($\omega=2/5/20$ op/voxel) and normalized to baselines. The respective baselines were computed from the same workloads performed using 16 cores only. The macaque dMRI dataset was chosen for this experiment because it is the largest dataset that can be processed on a single computing node of the Irène-Rome cluster, allowing us to evaluate strong scaling without introducing additional overhead from inter-node communication. The strong scaling demonstrates the benefits of parallelization, achieving reduced computation time by effectively reducing the workload per processor without introducing excessive overhead.

The weak scaling curve was established using the post-mortem human brain Chenonceau dataset to illustrate scaling behavior on an artificially growing dataset while preserving the workload per processor. To this aim, the tractography domain of the Chenonceau brain was split into 800 parcels. From 128 cores to 12800 cores, corresponding to 1 and 100 nodes, respectively, 8 to 800 parcels were involved, using 16 cores per MPI process. Similar to the strong scaling curve, the speed-up was normalized to the time required for the same workload performed on a single parcel.

\begin{figure}[H]
\centering
\includegraphics[width=0.85\columnwidth]{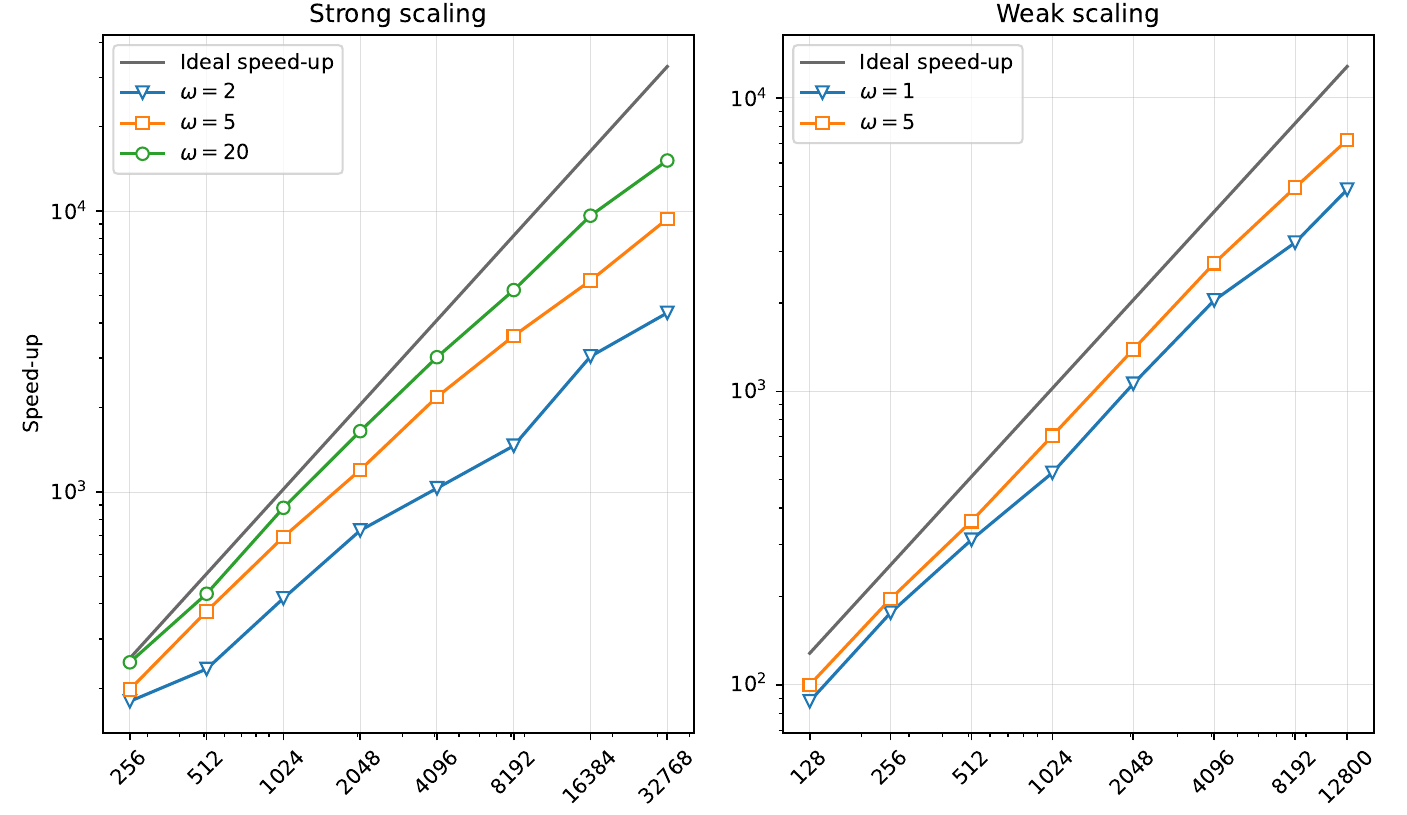}
%% Use \caption command for figure caption and label.
\caption{Scalability curves of ExaTract on the Irène-Rome CPU partition. The speed-up was evaluated in strong (left) and weak (right) scaling problems using from 1 node (128 cores) up to 256 nodes of the Irène-Rome partition.
}\label{figScalability}
\end{figure}

As expected, the workload plays an important role in the performance. While a heavy workload improves computational performance, a low workload requires more communication to reduce errors. Both experiments show correct scalability (Figure \ref{figScalability}). While the strong scaling curve shows significant acceleration on a given dataset, the weak scaling curve demonstrates ExaTract's ability to handle very high-resolution, large datasets that require substantial memory. The linear asymptotic behavior of the weak scaling curve suggests that scalability is even feasible for larger datasets. This property is strategically aligned with the emergence of very large datasets in microscopy, combining dMRI, 3D-PLI, or comSLI, such as the BigMac dataset \citep{howard}. Thanks to the assumptions made, especially the domain decomposition, an MPI process dealing with a given brain parcel interacts only with the processes in charge of neighboring parcels, ignoring the rest. It means upscaling depends only on the available resources for the largest tractography domains, which supports high scalability potential on future exascale supercomputers.

\subsection{Proof of scalability for population imaging}

\begin{figure}[H]
  \centering
  \rotatebox{0}{\includegraphics[width=1.0\columnwidth]{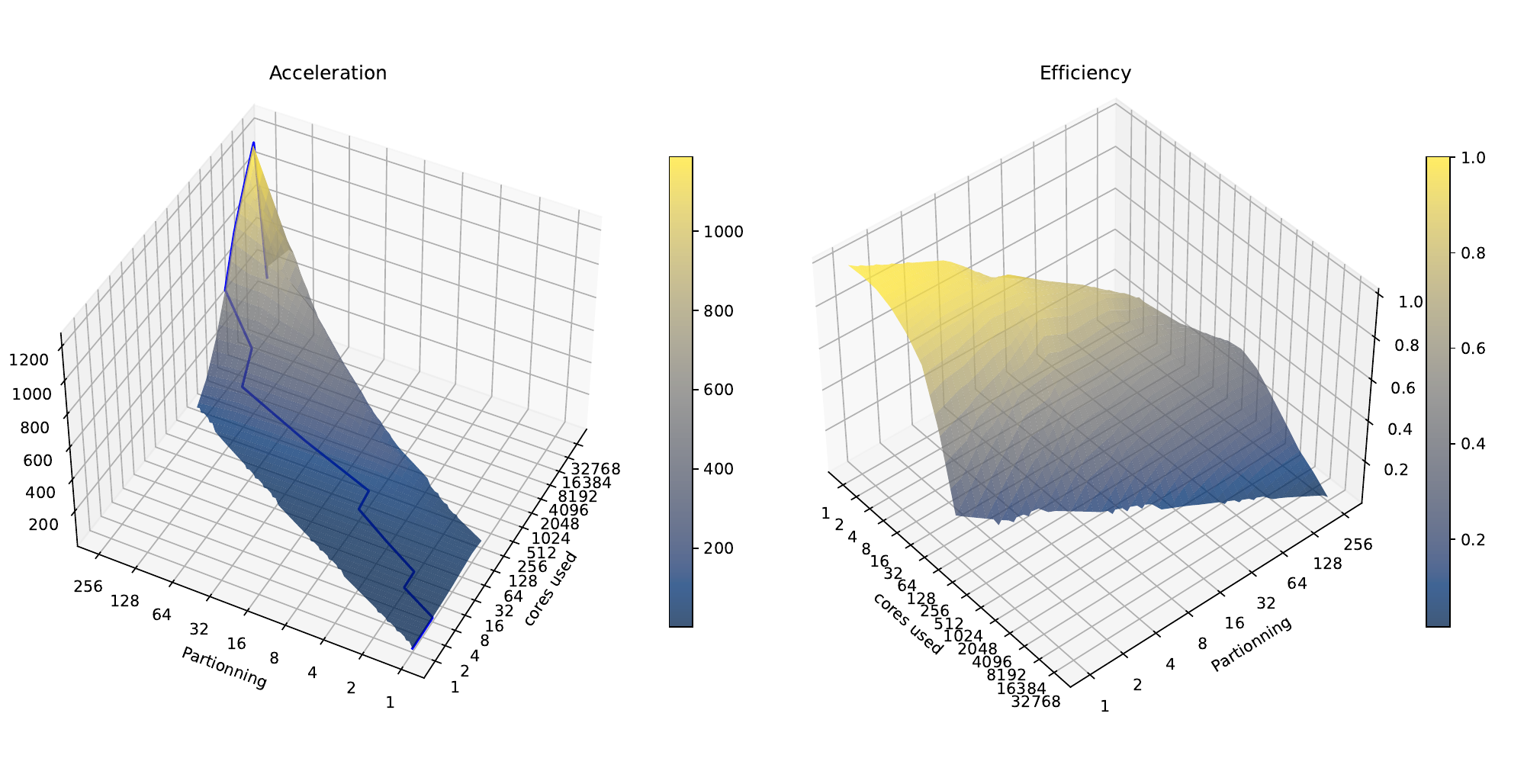}}
  \caption{Left: Speed-up landscape across different deployment configurations on the Irène-Rome CPU partition. The speed-up is calculated using the reference time obtained with one MPI process launched on one core. The blue curve illustrates the optimal configuration that provides the fastest tractography at a fixed number of cores. Right: efficiency across different deployment configurations. The reference time stemming from a deployment on one MPI process and 1 core was used to compute the efficiency.}
  \label{fig:chap_6acc}
\end{figure}

Before testing ExaTract against the entire HCP dataset, we benchmarked the performance of the designed parallelized tractography algorithm on the available hardware by mapping the performance landscape across various deployments on the Irène-Rome computer. The tractography domain of one HCP subject was partitioned into powers of 2 from 1 to 256. For each partitioning, the number of threads per MPI process and the required number of nodes (up to one partition per node) were varied. Different tractography experiments were run with an evaluation of the first 10 MPI iterations' performance. The speed-up and corresponding efficiency are respectively shown in Figure~\ref{fig:chap_6acc}.

\begin{figure}[H]
  \centering
  \rotatebox{0}{\includegraphics[width=0.85\columnwidth]{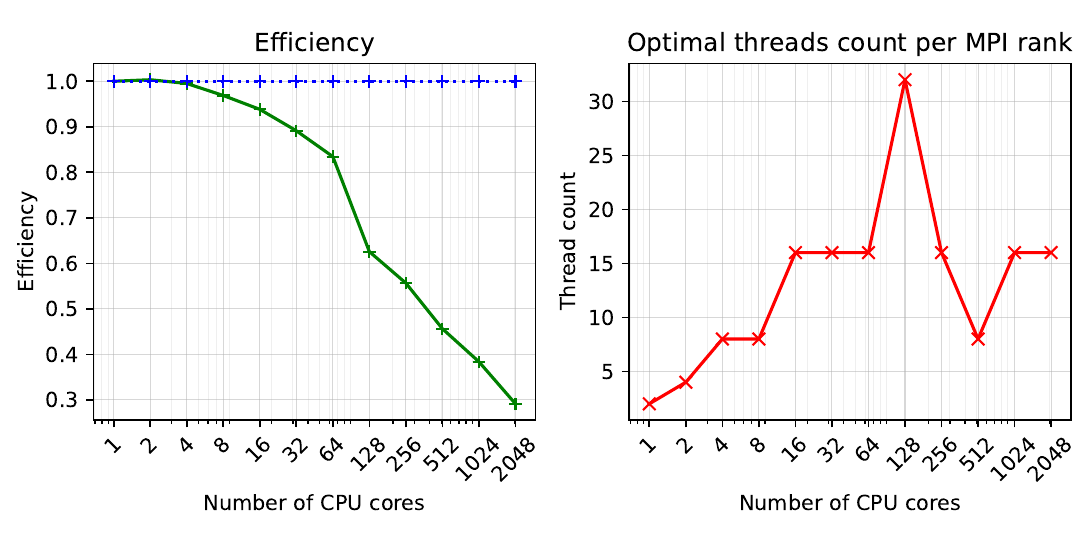}}
  \caption{Left: profile of the optimal efficiency obtained among the tested deployments on an HCP subject. Right: corresponding optimal number of threads used per MPI process (or per partition). Over 16 cores used, the optimal number of threads used per MPI process never exceeds 32, illustrating the need to split the global tractography task into different MPI processes inside a node.} %TODO
  \label{fig:chap6_optimalConfig}
\end{figure}

The optimal settings that yield the best performance at fixed numbers of cores are represented by the blue curve in Figure~\ref{fig:chap_6acc}. Up to 2048 cores (corresponding to 16 nodes), the performance profile, corresponding to these optimal settings (blue curve), is also reported in Figure~\ref{fig:chap6_optimalConfig}.

At millimeter resolutions, the scalability up to 32,768 cores (or 256 nodes) is, as expected, nonsensical. The experiment was primarily conducted for ExaTract behavior characterization. The efficiency remains above 50\% up to 256 cores (2 nodes), whereafter it drops below this limit. Interestingly, the optimal configurations, as displayed in Figure~\ref{fig:chap_6acc} and Figure~\ref{fig:chap6_optimalConfig}, show that the optimal number of threads per MPI process is never the absolute maximum of 256 but varies between 8 and 32. This is due to the non-uniform memory accesses of the NUMA nodes of the Irène-Rome computers. Memory access times, therefore, depend on the thread's position and the location of the allocated memory. It plays in favor of the use of more MPI processes to better benefit from the use of memory proximity tailored to the hardware. For this reason, the number of threads per MPI process was fixed at 32, corresponding to 16 cores, for the following experiments on the HCP and Chenonceau datasets.

\begin{figure}[H]
\centering
\includegraphics[width=1.0\columnwidth]{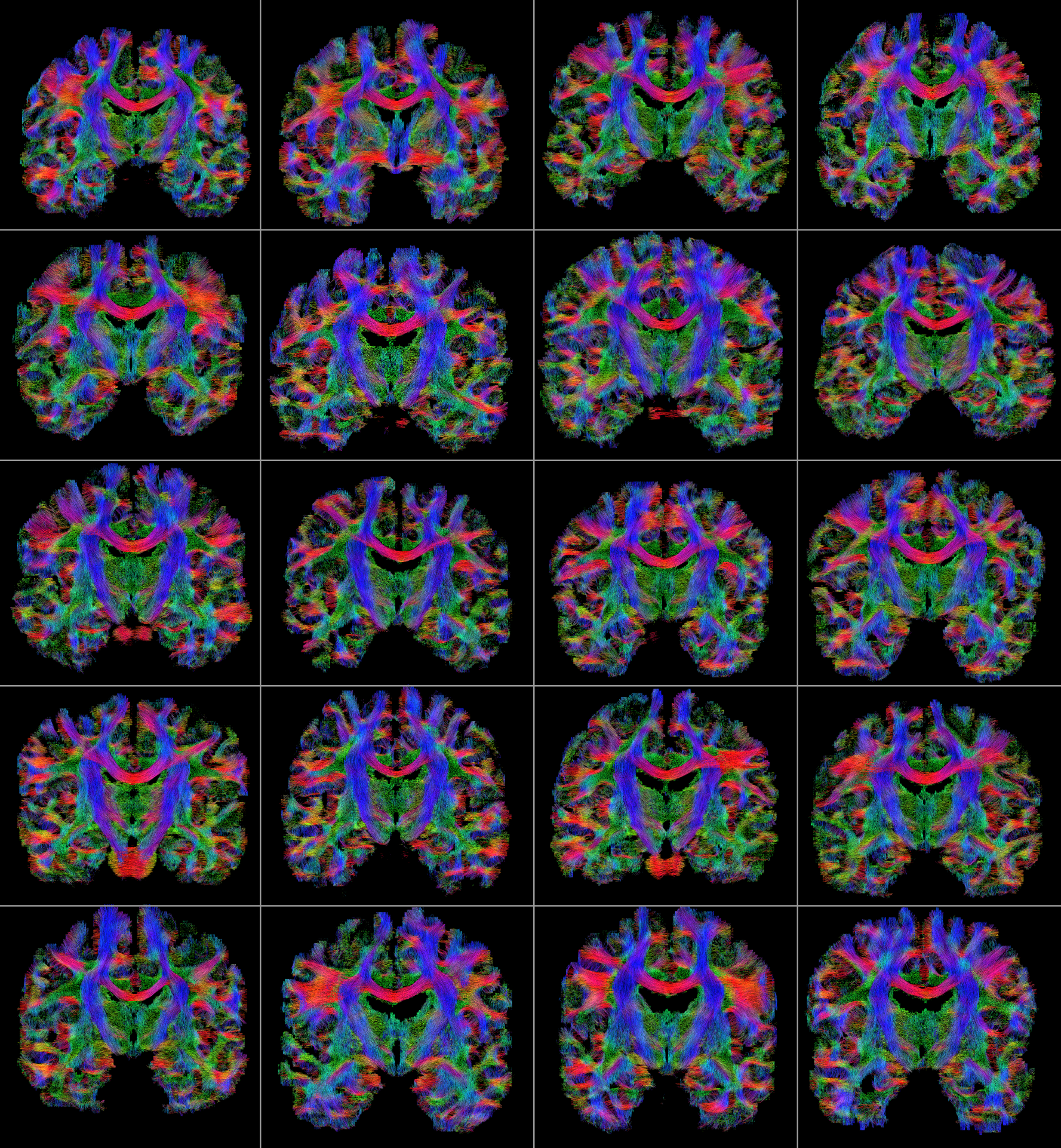}
\caption{Coronal views of 20 randomly selected tractograms in their native space computed with ExaTract among the 1065 subjects of the HCP cohort.}
\label{entireHCP}
\end{figure}

\begin{figure}[H]
\centering
\includegraphics[width=0.70\columnwidth]{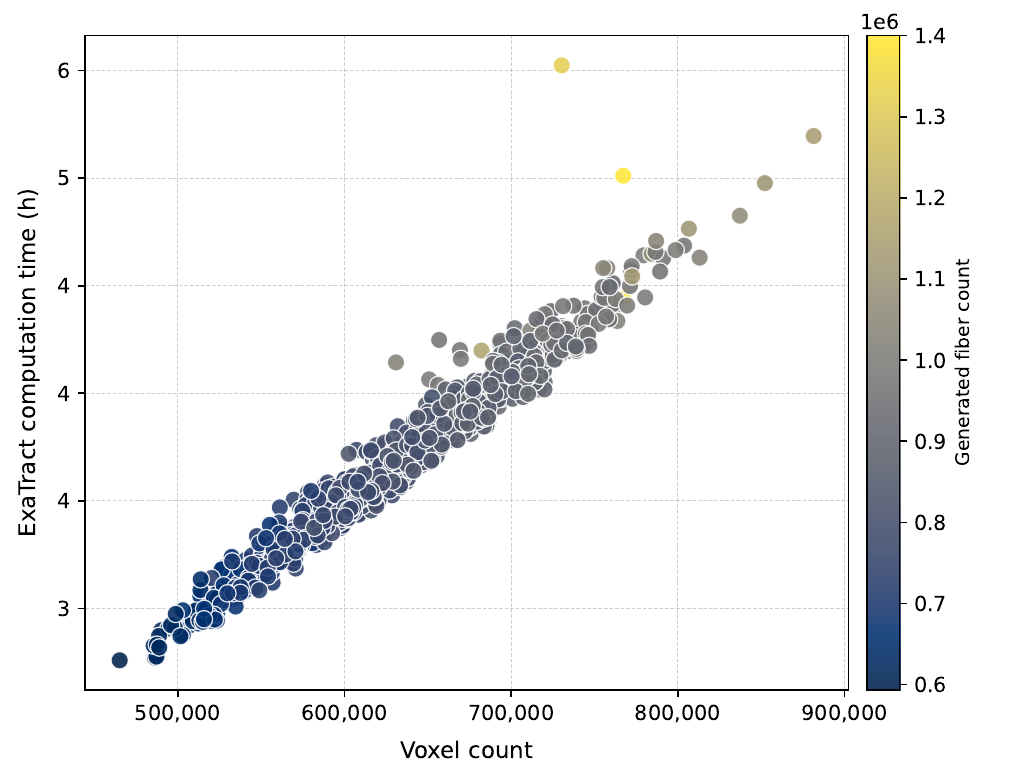}
%% Use \caption command for figure caption and label.
\caption{Global tractography Computation time with ExaTract for each HCP subject with respect to voxel count in the tractography domain, colored by the number of generated streamlines. The plot shows the expected linear relationship between the two variables. The voxel count directly determines the number of operations, and with a similar density of spin-glasses per voxel, the computation time increases linearly with the tractography domain size.}\label{hcpStat}
\end{figure}

Due to the Irène-Rome supercomputer scheduler's preference for whole-node usage, we constrained the ExaTract deployment to a single HCP subject per NUMA node. In that case, the optimal configuration for 128 cores is to use 8 MPI processes, each with 32 threads. Using 178 nodes of the Irène-Rome HPC facility simultaneously (totaling 22784 CPU cores), the application of the ExaTract tractography method over the 1065 subjects took less than 24 hours. With $8.07\text{x}10^5$ streamlines generated on average per subject in 3 h 40 min, the obtained tractograms are at least as comprehensive in terms of fiber count as those generated by the conventional streamline algorithm using one seed per voxel to create a streamline. The correspondence between the computation time per subject and its tractography domain size is shown in Figure~\ref{hcpStat}. In total, $8.59\text{x}10^8$ fibers were generated across the 1065 subjects in 22 hours. This makes it the largest global tractography dataset ever computed, in terms of fiber count. A small sample of these results is displayed in Figure~\ref{entireHCP}.

\subsection{Proof of scalability on an ultra-high resolution dataset}

ExaTract is not only a valuable tool for population imaging analysis; it is also of major interest in the community to deal with ultra-high mesoscopic resolution datasets. In this second application, we focused on the use of ExaTract to infer the structural connectivity of the Chenonceau dataset, an ultra-high-field diffusion MRI dataset of a post-mortem human brain scanned at the mesoscale.

\begin{figure}[H]
\centering
\includegraphics[width=1.0\textwidth]{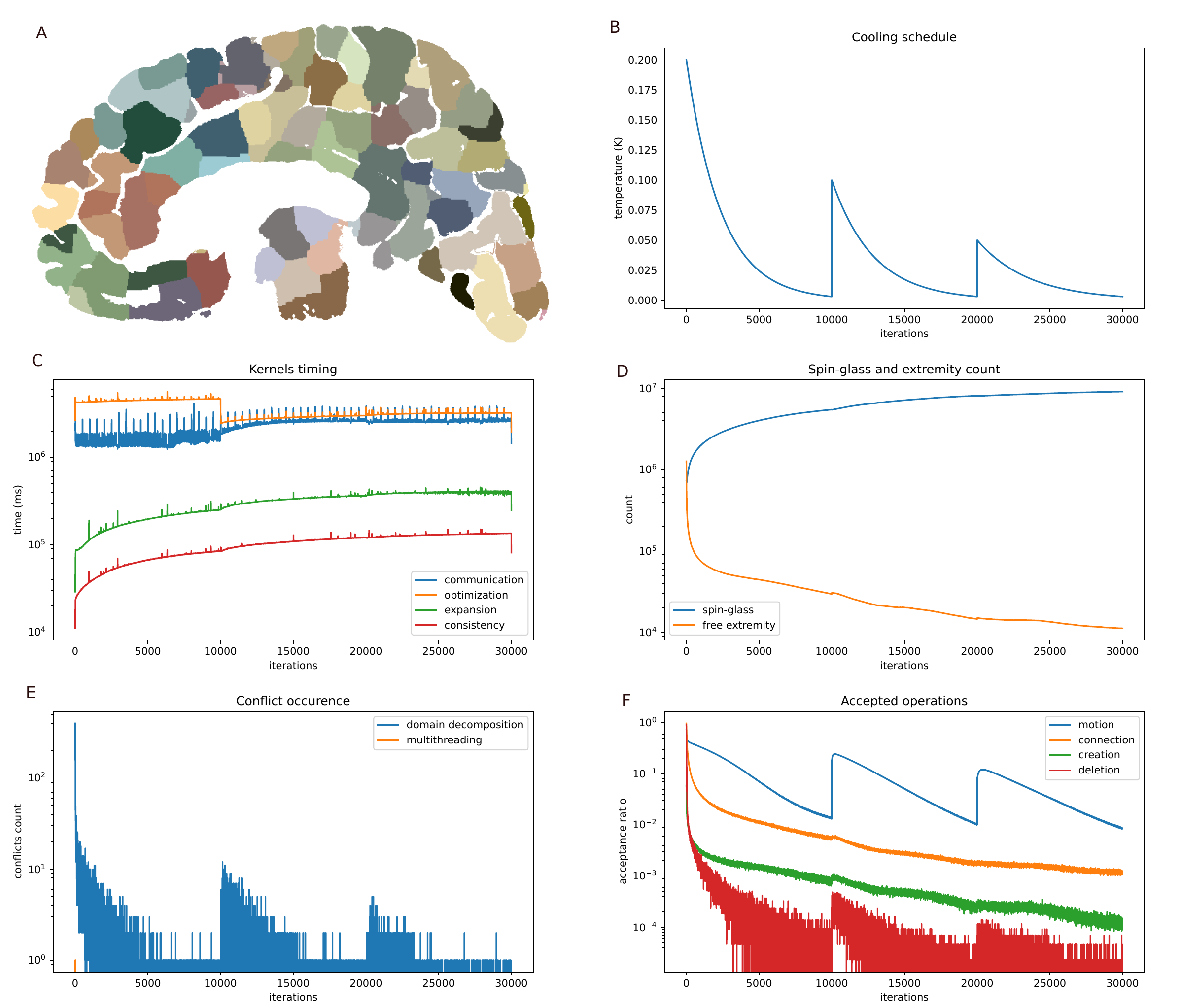}
\caption{Application of ExaTract to the Chenonceau dataset. (A) The entire brain was partitioned into 800 parcels using METIS \citep{karypis_metis_1997}. (B) The cooling schedule was tuned to ensure convergence over three runs, with a logarithmic temperature decrease. (C-F) Timing and convergence analysis performed on rank 0. (C) Kernel timings across ExaTract iterations reveal the dominance of optimization kernels over expansion and consistency kernels. Communication overhead remains comparable to the optimization cost, which is an expected trade-off when scaling to datasets of this size. (D) The evolution of spin-glass count and free extremities across the three runs directly confirms the system's proper convergence. (E) Conflict occurrence remains negligible relative to the total number of operations. (F) The evolution of the accepted modification ratio across operation types further illustrates the system's reconfiguration dynamics during the convergence.}
\label{chenonceauStat}
\end{figure}

The tractography domain of the Chenonceau brain was partitioned into 800 parcels (see Figure \ref{chenonceauStat}A) that were distributed over 100 nodes of the Irène-Rome CPU partition, corresponding to 12800 bi-threaded CPU cores. Given the previous benchmark of section~\ref{benchmarkParagraph}, the efficiency of this deployment is expected to be around 0.6 at $\omega = 5$ op/voxel compared to a 32-threaded approach, which remains a good trade-off between workload and communication. The number of threads per MPI process was set to 32 to favor local memory usage.

Three runs were performed to reach convergence, with a gradual increase in the workload between runs and a decrease in the cooling schedule to reach convergence. At the end of each run, temperature schedules were empirically adapted to keep the convergence tendency. The three simulated annealing cycles comprised 10,000 MPI iterations each with an exponential decrease from 0.2 to 0.003 K for the first cycle, from 0.1 to 0.003 K for the second cycle, and from 0.05 to 0.003 K for the third cycle (see Figure \ref{chenonceauStat}B). Initial spin-glass per voxel was set to $n_{init}=4$ spin-glass/voxel and $l=r_{search}=160$ $\mu$m. To speed up convergence, the number of sampled directions in the ODFs was reduced to 1000, and $\omega$ was increased from 5 to 10 op/voxel for the second and third runs. In Figure \ref{chenonceauStat}C-F, the timing and convergence analysis of the three runs are displayed. The optimization kernels, which represent the core of the tractography process, are predominant compared to the expansion and consistency kernels. Communication overhead remains acceptable compared to the optimization time and in agreement with the previous benchmark. It is worth noting that the regular peaks observed in the communication kernel correspond to intermediate saves of the spin-glass map to the storage disk (every 500 iterations). The ratio of accepted random motion is highly dependent on the system temperature, allowing for reconfigurations and exploration by altering spin-glass positions and orientations. The evolution of the accepted modification ratio for creation, deletion, and connection operations illustrates the system's expected convergence, with a continuous decrease in acceptance rates throughout the progression of the tractography. Similarly, the number of spin-glasses and free extremities across the three runs correlates with the expected behavior, with a monotone decrease of free extremities and a logarithmic growth of the number of spin-glasses reaching an asymptote, reflecting the direct effects of the connection, deletion, and creation operations. Conflict occurrence remains very low relative to the total number of operations and spin-glasses, validating the proposed parallelization scheme.

\begin{figure}[H]
\centering
\includegraphics[width=1.0\textwidth]{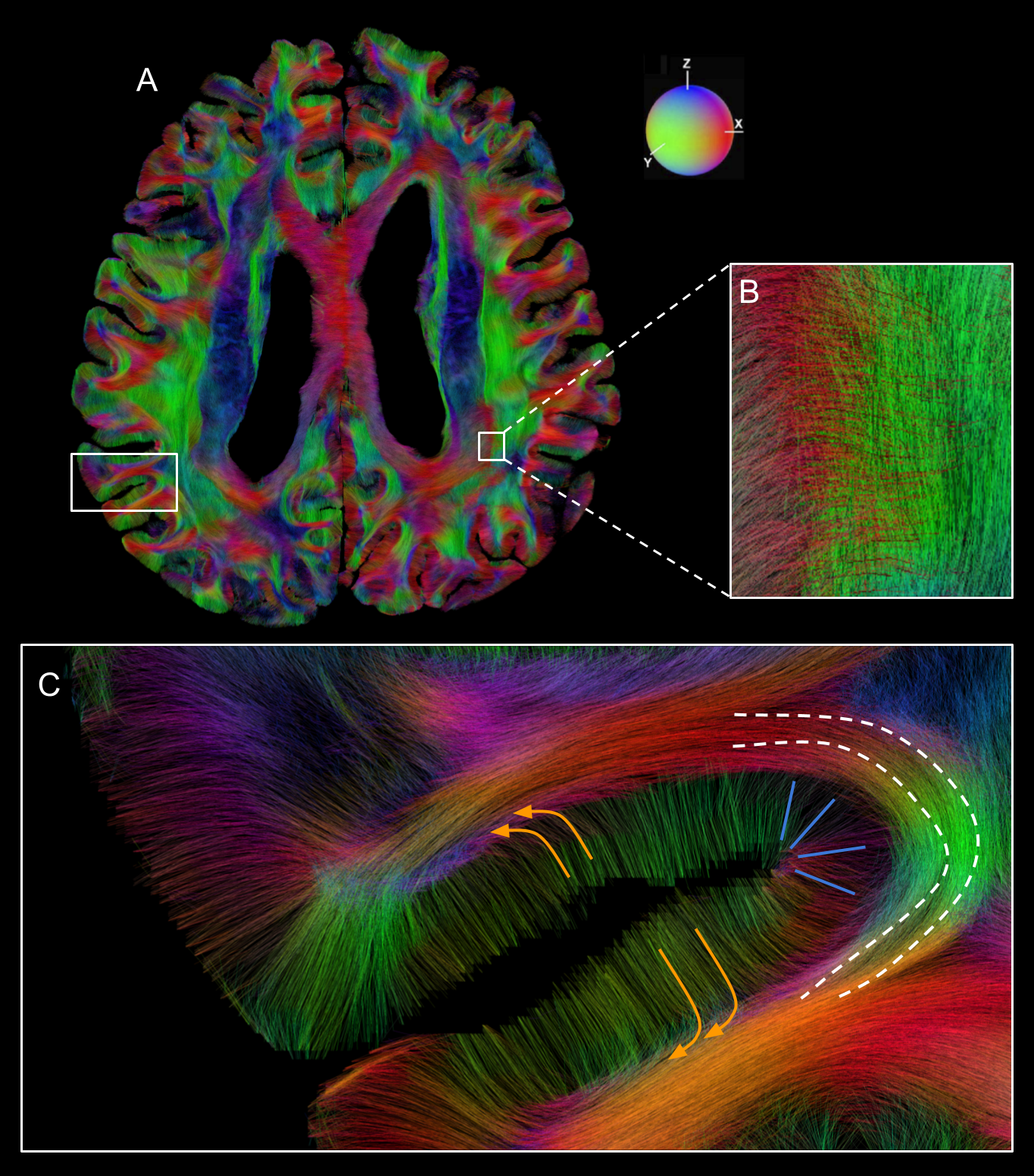}
\caption{Details of the global tractography results obtained on the Chenonceau brain in its native space. The entire tractogram yields approximately 25 million fibers; here, 1/4 of the total fibers are displayed in the axial view (A). Fibers in a crossing area (B) show the ability of the global tractography to recover complex fiber configurations, such as crossing. When zooming on a sulcus (C), superficial WM bundles joining two gyri are immediately visible (white dotted lines). Deep in the sulcus, fibers in the gray matter keep a radial trajectory (blue lines). Interestingly, some fibers originating in the cortex run along the gyrus (orange arrows), indicating even shorter connections.}
\label{chenonceauTracto}
\end{figure}

\begin{figure}[H]
\centering
\includegraphics[width=1.0\textwidth]{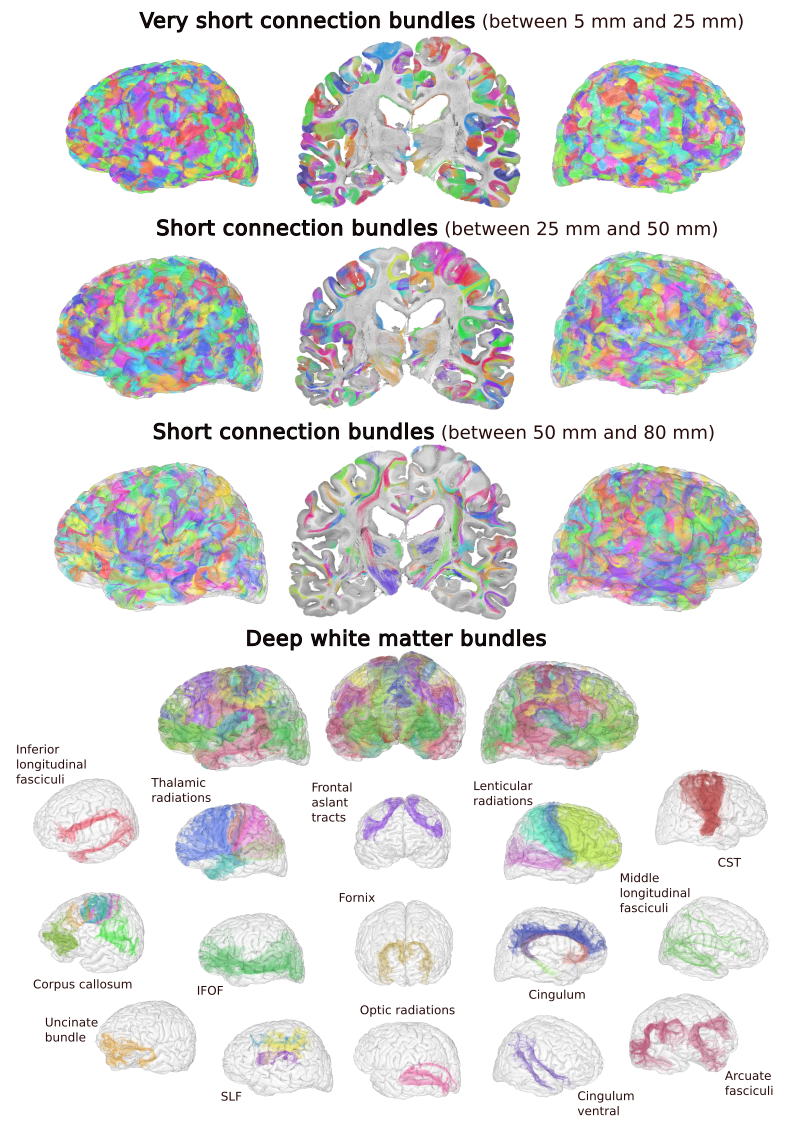}
\caption{Global tractography atlas of the Chenonceau brain. Fiber bundles stemming from a clustering of the tractogram (see \citet{legeay}) were sorted per length to expose the different connectivity levels. Very short connections, ranging here from 5 to 25 mm, remain particularly superficial, or even intra-cortical, while longer connections extend deeper into the white matter. The established deep WM bundles were extracted using an automatic procedure based on the atlas of \citet{herlin_sex-related_2024}. It includes the inferior longitudinal fasciculi, the thalamic radiations, split into the temporal (turquoise blue), frontal (blue), central (orange), parietal (pink) and occipital (green) thalamic radiations, the frontal aslant tracts, the frontal (green), central (blue), parietal (turquoise blue), pink (red) lenticular radiations, the cortico-spinal tracts (CST), the corpus callosum, the inferior fronto-occipital fasciculi (IFOF), the fornix, the cingulum bundles, split into the dorsal, long (red) and short (blue), and dorso-ventral (green), the middle longitudinal fasciculi, the uncinate bundles, the SLF, including the SLF I (blue), II (yellow) and III (purple), the optic radiations, the cingulum ventral, the arcuate bundles.}
\label{chenonceauBundle}
\end{figure}

It respectively took 18.6, 16.3, and 17.9 hours for the three runs to be completed, totaling 52.8 hours of consecutive computation. At a machine level, it represents 675,840 core hours of cumulative computation. A total of $4.2\text{x}10^9$ spin-glasses were involved in the $1.74\text{x}10^{13}$ distributed operations, generating $25\text{x}10^6$ fibers. Figure \ref{chenonceauTracto} illustrates the obtained tractogram computed from ExaTract, with an overview of anatomical details obtained thanks to the combination of the global tractography algorithm with the high-spatial and angular resolutions of the Chenonceau dataset. To further assess the correct convergence of the tractography, the deep WM bundles were automatically labeled similarly to the methods developed in section~\ref{labelling}. The obtained bundles are displayed in Figure \ref{chenonceauBundle} and show a good recovery of the well-established deep WM bundles, such as the arcuate fasciculus, the inferior fronto-occipital fasciculus, the inferior longitudinal fasciculus, the uncinate fasciculus, the cingulum bundle, the corpus callosum, and the corticospinal tract.

As presented in \citet{legeay}, the tractogram was further processed with clustering methods to extract fiber bundles. Figure~\ref{chenonceauBundle} depicts all bundles according to their length range. The amount of found bundles using global tractography exposes the richness of the short connectivity in the human brain, from very short intra-cortical connections up to deeper bundles connecting adjacent gyri. %TODO ajouter plages de longueur

\section{Discussion}

Thanks to spatial decomposition and an efficient multithreading strategy, ExaTract effectively reduces computation time while allowing more fibers to be generated. Through dedicated experiments on the DiSCo phantom and an HCP subject, we demonstrated that the parallelization scheme does not significantly affect the system's convergence, even in extreme domain-decomposition cases. The obtained tractograms remain similar across different parallelization strategies, and the conflict occurrences remain negligible compared to the total number of operations performed. It remains, however, challenging to further characterize convergence in overlapping areas, as conflicts may have additional effects on the system's convergence. While the experiments do not show significant differences in the resulting tractograms, it is possible that convergence is locally and dynamically affected in overlapping areas, which may not be visible at the global level or at the end of the convergence process. Future work could focus on further characterizing the convergence in these areas dynamically, yet, due to the randomness of the MCMC process and the many possible variations in the energy landscape, these effects might be difficult to characterize and may require a very large number of experiments to be statistically significant.

The scalability benchmarks further illustrate ExaTract's ability to effectively reduce computation time while maintaining high efficiency, even at large numbers of cores. In comparison, tools like MRtrix \citep{reisert_global_2011,christiaens_2015} or MITK \citep{neher} face time and memory constraints that limit the number of generated fibers (typically < 100,000) and the resolution of the datasets they can handle. This can fall short when representing the structural connectivity of the entire human brain. We reported in Table~\ref{comparisonTable} the performance of different global tractography tools on human datasets. While the comparison is difficult due to differences in the experimental setup, it provides an order-of-magnitude estimate of ExaTract's performance relative to existing tools. In a similar computational time, ExaTract generates 10 times more fibers than the tool from the previous implementation, while optimizing 10 more spin-glasses in a slightly more resolved dataset. As shown earlier, with 4096 threads, ExaTract achieved the same fiber-generation performance as the 256-thread implementation in less than 12 minutes, making it suitable for clinical research investigations.

It is worth noting that the comparisons remain highly sensitive to the experimental setup: first, computing hardware design technologies have advanced substantially since the first demonstration of global tractography. Second, the chosen initial conditions of the global tractography algorithm can significantly impact the global tractography process. The initial spin-glass density per voxel directly influences the total number of generated streamlines and increases the workload when searching for potential connections. Other parameters linked to the diffusion model, such as the number of directions used to sample the ODF or the choice of a generative spin-glass model, directly affect the workload. Finally, spin-glass length and research diameter can affect the convergence speed.

In addition to time constraints, current implementations cannot handle very large datasets, such as the Chenonceau dataset, due to memory limitations. ExaTract, on the other hand, is designed to handle such large datasets by distributing the workload across multiple nodes and effectively managing memory usage through domain decomposition. This allows ExaTract to perform global tractography on ultra-high-resolution datasets that would be infeasible with state-of-the-art tools. An additional advantage of ExaTract is its available memory obtainted through spatial decomposition. Computation at the level of the diffusion model used is often demanding due to the limited memory available in previous implementations of tractography. With additional memory, ExaTract could keep the orientation probabilities in RAM for each voxel, which are currently computed on the fly. This would allow us to save a lot of time by avoiding the repeated computation of the diffusion model for each spin-glass and each operation.

\begin{table}[H]
\centering
\begin{tabular}{l | r r r r r }%% Table column specifiers
%% Tabular cells are separated by &
  \hline
  Global tractography & resolution & spin-glass & fiber & computation & threads \\
  implementation & ($mm^3$)& count & count & time & used \\
  \hline
  \citet{fillard_2009}                  & 1.8x1.8x2      & $2\text{x}10^6$ & - & 3 days & - \\
  \citet{reisert_global_2009}      & 2x2x2          & $2.5\text{x}10^6$ & $10^5$ & <24h & - \\
  \citet{wu_embarrassingly_2016}     & 2x2x2          & $2.5\text{x}10^6$ & $2\text{x}10^5$ & ~1h30min & 10 \\
  \citet{konopleva_modelfree_2018} & 1.5x1.5x3      & $3\text{x}10^6$ & - & 6h & 2 \\
  ExaTract (HCP)                  & 1.25x1.25x1.25 & $3\text{x}10^7$ & $8\text{x}10^5$ & 2h40min & 256 \\
  ExaTract (HCP)                  & 1.25x1.25x1.25 & $3\text{x}10^7$ & $8\text{x}10^5$ & \textbf{11min48s} & 4096 \\
  ExaTract (Chenonceau)           & 0.2x0.2x0.2    & $4.2\text{x}10^9$ & $2.5\text{x}10^7$ & 52h48min & 12800 \\
  \hline
\end{tabular}
\caption{Comparison of fiber count and convergence timings for multiple tractography implementations in human datasets.}\label{comparisonTable}
\label{comparison}
\end{table}

A strong limitation of global tractography is the difficulty in predicting the convergence time. The number of generated fibers and spin-glasses, as well as the convergence speed, are highly dependent on the initial conditions of the tractography. This makes it challenging to estimate the computational resources and time required for a given dataset, especially for large, high-resolution datasets. The additional parameters involved in the parallelization add complexity to setting up the tractography. Users need to find the right balance between workload and communication, which can be difficult without prior experience or benchmarks. This can lead to suboptimal performance if not properly configured. Future work could focus on developing more adaptive parallelization strategies that dynamically adjust the workload and communication based on dataset characteristics and available computational resources, potentially improving efficiency and reducing the time required for global tractography. One of the key advantages of the portability library Kokkos used in ExaTract is its capability to support emerging hardware architectures with minimal code modifications, from CPU to GPU-based architectures.

While a non-generative approach for the spin-glass model was implemented in this work, ExaTract is designed to be flexible and adaptable to various energy models, opening the way to plug any local model of the diffusion process and any regularization term to improve the robustness of the tractography. This flexibility allows ExaTract to be used with different global tractography frameworks. However, generative approaches, as first introduced by \citet{kreher_gibbs_2008}, involve a modeling of the contribution of each spin-glass in the diffusion-weighted signal. The optimization process seeks to minimize the voxel-wise difference between the generated diffusion signal from spin-glasses and the actual diffusion data by expressing the external energy as follows:

\begin{equation}
    E_{\text {ext }}(\mathscr{M}, D)= \sum_{\text{voxel} \; \mathbf{V}} \left\| \sum_{X_i \in \mathbf{V}} \rho_{X_i} - D_{\mathbf{V}} \right\|^2 \; \text{,}
\end{equation}

\noindent where $\rho_{X_i}$ is the signal generated by the spin-glass $X_i$ according to a modelling assumption and $D_\mathbf{V}$ is the measured diffusion data in voxel $\mathbf{V}$. \citet{reisert_mesoft_2014}, \citet{christiaens_2015} and \citet{konopleva_modelfree_2018} have recently unified diffusion modeling and global tractography. Under the name of model-free global tractography, \citet{konopleva_modelfree_2018} makes the minimal assumption possible on the fiber response function of a spin-glass, meaning that the fiber response is re-estimated per voxel after each modification of the spin-glass configuration by an optimization under constraint over the fiber response function. In these cases, the external energy is computed voxel-wise, with all spin-glasses in each voxel taken into account to estimate the model. This approach requires some adaptation of ExaTract's parallelization scheme, since conflict occurrences are likely to increase when performing the reduction at the voxel level. A possible adaptation of the model-free global tractography to ExaTract could be to perform the reduction at the iteration level, meaning that the fiber response function is updated at the end of each MPI iteration, thereby allowing the parallelization scheme to remain unchanged.

\section{Conclusion}

We demonstrate the scalability and performance of ExaTract on large datasets, showcasing its ability to reconstruct brain connectivity at unprecedented scales while maintaining accuracy. The entire HCP dataset was processed in under 24 hours using ExaTract, demonstrating its capability to handle large-scale neuroimaging data. The application of ExaTract to the ultra-high-resolution Chenonceau dataset further illustrates its ability to reconstruct detailed brain connectivity patterns at a mesoscopic scale. This achievement allowed us to build the first mesoscopic structural connectome from global tractography. By addressing the computational challenges associated with global tractography, ExaTract opens new avenues for exploring the advantages of global approaches in both research and clinical settings. ExaTract is completely open-source and available on Framagit as part of the ginkgo toolkit.

\section*{Acknowledgments}

This research was supported by the Joint German-French Institute AIDAS Cycle 1 and Cycle 2, by the European Union’s Horizon 2020 Framework Program for Research and Innovation under the Specific Grant Agreement No. 945539 (Human Brain Project SGA2 and SGA3), and by the Priority Research and Equipment Program (PEPR) Santé Numérique BrainDeepPhenotyping. All the acquisitions performed for the Chenonceau dataset were entirely covered by the NeuroSpin imaging platform.

This project was provided with computing HPC and storage resources by GENCI at TGCC, thanks to the grants A0120310851 and SS010315366 on the supercomputer Joliot Curie's ROME partition. The authors would also like to thank Dr. France Boillod-Cerneux and Dr. Christophe Calvin for their support within the framework of the French-German AIDAS institute, as well as the GENCI French national HPC infrastructure.

Data were provided in part by the Human Connectome Project, WU-Minn Consortium (Principal Investigators: David Van Essen and Kamil Ugurbil; 1U54MH091657), funded by the 16 NIH Institutes and Centers that support the NIH Blueprint for Neuroscience Research; and by the McDonnell Center for Systems Neuroscience at Washington University.

\appendix
\section{Conflict occurence estimation}
\label{app1}

In the case of parallelization on a shared-memory architecture, the probability of conflict occurrence can be estimated as follows. During an ExaTract iteration, the total workload performed on a partition with $N_{v}$ voxels, with $n$ spin-glasses each, is divided by all $N_t$ working threads, so that each thread performs $\frac{N_v\omega}{N_t}$ operations per iteration. For a given spin-glass, the probability to be drawn by a thread is $1-(1-\frac{1}{N})^{\frac{N_v\omega}{N_t}}$. Supposing that all operations have a duration $t$ and that the beginning of operations on two threads are independent and uniformly distributed in $[0, t]$, the probability of overlapping modifications for two threads is $1 - (1 - \frac{t_c}{t})^2$, where $t_c$ is the critical zone duration. Hence, the probability of multithreading conflict occurrence $X$ for a given spin-glass can be estimated as follows:

\begin{equation}
    p(X) = \left(1- \left(1-\frac{1}{nN_v}\right)^{\frac{N_v\omega}{N_t}}\right)^2 \left(1 - \left( 1 - \frac{t_c}{t} \right)^2 \right) \overline{R}^2
\end{equation}

Given the $\binom{N_{t}}{2}$ different pairs of threads that can potentially modify the same spin-glass, and according to the linearity of the expectation over the entire spin-glass set, the mean conflict occurrence during an iteration, normalized per operation, is given by:

\begin{equation}
    \frac{\mathbb{E}[X]}{\omega N_v} = \binom{N_{t}}{2} \frac{n}{\omega} p(X)
    \label{multithreading_conflict}
\end{equation}

Supposing that modification draws are independent and uniformly distributed per voxel, and that a modification made in the overlapping area of a partition has a probability of $p_{overlap}$ to alter a spin-glass in a neighboring partition, the expectation of communication conflict occurrence $Y$ in an overlapping area can be estimated as follows:

\begin{equation}
    \frac{\mathbb{E}[Y]}{\omega N_v} = \frac{n}{\omega} \left(1- \left(1-\frac{1}{n}\right)^{\omega}\right)^2 \overline{R}^2 p_{overlap}
    \label{communication_conflict}
\end{equation}

\section{Conflict resolution strategy}
\label{app2}

\begin{figure}[H]
\centering
\includegraphics[width=0.5\columnwidth]{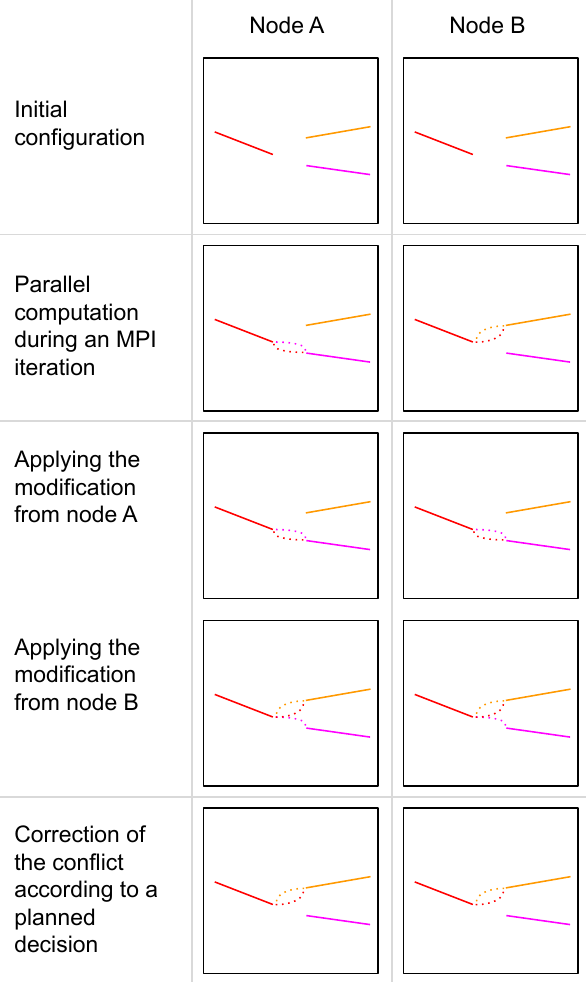}
\caption{Example of conflict and subsequent correction. Node A and node B start from the same initial configuration in a voxel shared in the overlapping area between the two nodes. While node A connects the red spin-glass to the purple one, node B independently connects the red spin-glass with the orange one. After communicating their respective modifications, the modification from node A is first applied to the previously stored initial configuration. The modification from node B is then applied, leaving the purple spin-glass non-reciprocally connected to the red spin-glass. This leads to an anomaly that is further detected and corrected according to a predefined resolution plan.}
\label{resolutionConflict}
\end{figure}

The approach for resolving conflicts is based on a kernel that detects inconsistencies in the spin-glass configurations and corrects them according to a predefined resolution plan. For this strategy to be effective, the merging of the modifications performed in the overlapping area was thought to be symmetrical and deterministic, ensuring that the same correction is consistently applied on all sides of the overlap without additional communication. This is achieved through a strategy of merging with \textit{a posteriori} corrections, which is divided into the following four steps:
\begin{enumerate}
    \item \textbf{Identification of modifications:} All modifications made within the overlap area are identified by comparing the current state with a saved copy of the overlap area before any modifications were performed.
    \item \textbf{Exchange of modifications:} Using the communication protocol, all these modifications are exchanged among neighboring nodes that share the same overlap area.
    \item \textbf{Application of modifications:} In a predefined order, arbitrarily chosen and planned in advance by all the nodes, the changes made by all the nodes sharing the overlap area are sequentially applied to the original copy of the overlap area. This process can lead to conflicts, as modifications could have been performed in parallel. Despite potential conflicts, all modifications are applied, potentially resulting in inconsistent spin-glass configurations, as illustrated in Figure \ref{resolutionConflict}.
    \item \textbf{Detection and correction of abnormalities:} Any abnormality, such as spin-glass inconsistencies within the overlap area, is detected and rectified according to a predefined resolution plan in which all possible configurations are anticipated. Figure \ref{resolutionConflict} exposes an example of a possible conflict resolution.
\end{enumerate}
This approach enables rapid recovery from conflicts, efficiently restoring the system to a consistent state with a linear complexity kernel.

\end{document}